\documentclass[acmtog,screen,nonacm]{dhara-preprint}
\makeatletter\@ACM@balancefalse\makeatother
\usepackage{amsmath,booktabs}
\usepackage{tabularx}
\usepackage{array}
\usepackage{enumitem}
\usepackage{tikz}
\usetikzlibrary{arrows.meta,positioning,fit,calc}
\usepackage{microtype}
\usepackage{graphicx}
\usepackage{url}
\usepackage{fvextra}
\usepackage{placeins}
\usepackage{dblfloatfix}


\setcopyright{none}
\acmDOI{}
\AtBeginDocument{\renewcommand{\keywordsname}{Keywords}}
\renewcommand\footnotetextcopyrightpermission[1]{}

\definecolor{KGGreen}{HTML}{228833}
\definecolor{ENBlue}{HTML}{0072B2}
\definecolor{GeminiOrange}{HTML}{EE7733}
\definecolor{SVGray}{HTML}{808080}
\title{Ontology-Guided Knowledge Graph Construction and Robust Evidence Retrieval for Indian Knowledge Systems}
\author{Anand Kaniyaar Sriranga Prasad}
\orcid{0009-0008-1360-7334}
\authornote{Anand Kaniyaar Sriranga Prasad and Avinash Holenarasipur Nagendra contributed equally and are joint first authors.}
\affiliation{\institution{HTSR Institute}\city{Sangareddy}\country{India}}
\affiliation{\department{Department of Heritage Science \& Technology}\institution{Indian Institute of Technology Hyderabad}\city{Sangareddy}\country{India}}
\author{Avinash Holenarasipur Nagendra}
\orcid{0009-0002-2800-5113}
\authornotemark[1]
\affiliation{\department{Department of Heritage Science \& Technology}\institution{Indian Institute of Technology Hyderabad}\city{Sangareddy}\country{India}}
\author{Mohan Raghavan}
\orcid{0000-0002-9196-8193}
\affiliation{\department{Department of Heritage Science \& Technology}\institution{Indian Institute of Technology Hyderabad}\city{Sangareddy}\country{India}}
\affiliation{\department{Department of Biomedical Engineering}\institution{Indian Institute of Technology Hyderabad}\city{Sangareddy}\country{India}}
\affiliation{\institution{Central Sanskrit University}\city{New Delhi}\country{India}}
\affiliation{\institution{Bheri Technology Foundation}\country{India}}
\renewcommand{\shortauthors}{Anand et al.}

\makeatletter
\newcommand{\dharaauthoraffiliations}[2]{%
  \ifcsdef{typeset@author#1}{%
    \csletcs{dhara@typeset@author#1}{typeset@author#1}%
  }{%
    \cslet{dhara@typeset@author#1}{\@firstofone}%
  }%
  \csdef{typeset@author#1}##1{%
    \csuse{dhara@typeset@author#1}{##1}\textsuperscript{#2}}%
}
\dharaauthoraffiliations{1}{2,1}
\dharaauthoraffiliations{2}{1}
\dharaauthoraffiliations{3}{1,3,4,5}

\def\dhara@skipaffiliation#1#2{%
  \let\and\@typeset@author@line}
\patchcmd{\@mkauthors@i}
  {\noindent\addresses\@typeset@author@line}
  {\noindent\let\affiliation\dhara@skipaffiliation\addresses\@typeset@author@line}
  {}%
  {\ClassError{dhara-paper}{Cannot apply shared-affiliation layout}%
    {Check the author layout against the installed acmart class.}}
\makeatother

\authorsaddresses{%
  Affiliations: \textsuperscript{1}Department of Heritage Science \& Technology,
  Indian Institute of Technology Hyderabad, Sangareddy, Telangana, India.\newline
  \textsuperscript{2}HTSR Institute, Sangareddy, Telangana, India.\newline
  \textsuperscript{3}Department of Biomedical Engineering,
  Indian Institute of Technology Hyderabad, Sangareddy, Telangana, India.\newline
  \textsuperscript{4}Central Sanskrit University, New Delhi, India.\newline
  \textsuperscript{5}Bheri Technology Foundation, India}

\newcommand{\dhara}{Dh\={a}r\={a}}
\newcommand{\dharabenchtitle}{Dhara Bench: An Evidence-Oriented Retrieval Benchmark for Indian Knowledge Systems}
\newcommand{\dharatitle}{Dhara: Evidence-Oriented Retrieval for Indian Knowledge Systems}
\date{}
\begin{abstract}
Retrieving evidence from Indian Knowledge Systems is complicated by transliteration variants, ambiguous names, and relations distributed across passages. We present an ontology-guided knowledge graph (KG) pipeline that links typed entities and relations extracted with Qwen3-32B to original textual evidence. Retrieval combines sense-aware grounding, bounded graph traversal, lexical anchors, relationship-vector search, and passage reranking. We evaluate KG alone, single-vector retrieval (SV), and KG unions with SV and Entity Notes on 320 questions from \dhara{} Bench, matching the per-question candidate count across internal systems under BGE and Qwen rerankers. Standalone KG has the highest Precision@5 under both rerankers and the highest MRR@20 in Counting, Composite, and Multi-hop among the evaluated internal configurations. With Qwen, KG achieves Multi-hop MRR@20 of 0.8722, compared with 0.7622 for the fully reranked KG + SV union. The KG + SV + Entity Notes union instead provides the highest overall Recall@20 (0.5134), compared with 0.4347 for KG. On the 15 Multi-hop questions, Qwen KG achieves complete-evidence MRR@20 of 0.4651 versus 0.3372 for SV, while both full unions complete ten questions within twenty results compared with nine for KG. Gemini 3.7 Flash, evaluated through manually reviewed source mappings, remains stronger on overall MRR and nDCG, while internal systems have category-specific advantages. All comparisons are descriptive. A separate deployment timing study measures mean retrieval times of 20.374 s for KG and 0.646 s for SV. The findings support selecting KG or its unions according to the required balance of early relevance, evidence completeness, and retrieval cost.
\end{abstract}
\keywords{knowledge graphs, Indian Knowledge Systems, ontology-guided extraction, evidence retrieval, entity resolution, retrieval augmentation}
\begin{document}
\raggedbottom
\maketitle
\pagestyle{empty}\thispagestyle{empty}
\hypersetup{pdfcreator={LaTeX}}

\section{Introduction}
Indian Knowledge Systems (IKS) and heritage encompass a continuing civilizational tradition of the Indian subcontinent, developed over millennia and connecting philosophy, scientific inquiry, ethics, and everyday life \citep{elements_iks}. This integrated perspective provides a conceptual foundation for understanding diverse knowledge practices rather than treating them as isolated disciplines. Their textual sources include narrative literature, philosophy, medicine, architecture, performance theory, and normative texts. Evidence retrieval across these traditions involves several distinct difficulties. A name can appear in multiple transliterations or through an epithet; the same surface form can refer to different entities; and a question can require connecting an entity, an action, and a contextual qualification rather than locating a passage with similar wording. Domain-specific terms also resist a uniform interpretation across genres. A useful retrieval system must identify relevant passages while retaining enough structure to distinguish these cases.

Dense retrieval provides a strong mechanism for discovering semantic similarity. Multi-vector representations can preserve more local matching information than a single pooled vector, as exemplified by late-interaction retrieval \citep{colbert}. However, similarity alone does not explicitly encode whether an entity performs an action, is the object of that action, belongs to a particular lineage, or participates in a relation in a specific textual context. A passage mentioning two entities is not necessarily evidence for a relation between them. Likewise, a highly similar passage may concern a different sense of a shared name. These are representational differences, not claims that dense retrieval is incapable of answering relational questions.

We investigate a complementary approach: construct a typed, source-linked graph and use it to discover evidence candidates before reranking the original text. The graph is an intermediate retrieval representation, not a substitute for textual evidence or an authoritative store of automatically verified facts. Its value lies in making entities, relations, and provenance available as explicit retrieval signals. Its weaknesses, including extraction omissions and incorrect merges, remain relevant to the final ranking.

This work is the third component of our research programme. The first, \emph{\dharabenchtitle}~\citep{dharabench}, establishes an expert-curated ground-truth dataset. The second, \emph{\dharatitle}~\citep{dhara}, studies evidence-retrieval methodologies including single-vector, multi-vector, and Entity Notes (EN) retrieval. The present paper focuses on KG construction and KG-based retrieval within that unified system. Multi-vector retrieval, Entity Notes retrieval and remapping, KG retrieval, and their combinations are all components of our system; Gemini is used only as an external comparison system.

Our contributions are fourfold. First, we describe a genre-conditioned extraction pipeline that combines explicit ontology specifications with Qwen3-32B and source-linked graph storage. Second, we develop a KG evidence-retrieval procedure integrating sense-aware resolution, local graph structure, lexical anchors, and relationship-vector search. Third, we compare KG alone, SV alone, and their unions with mapped Entity Notes under matched per-question candidate budgets and two rerankers, including a comparison of full reranking and intersection pinning. Fourth, we assess category-specific relevance and complete-evidence retrieval using the metrics of \dhara{} Bench~\citep{dharabench}, compare against the reviewed Gemini reference from \dhara{}~\citep{dhara}, and characterize retrieval and reranking time in a separate deployment study.

The central finding is a distinction between early relevance and evidence coverage. Standalone KG achieves the highest MRR@20 in Counting, Composite, and Multi-hop among the evaluated internal configurations under both rerankers, whereas unions generally recover more annotated evidence within twenty results. With Qwen, the KG + SV + EN union has the highest overall MRR@20, Recall@20, and nDCG@10, while KG retains the highest Precision@5. Complete-evidence scoring similarly distinguishes earlier completion from completing more questions: Qwen KG has the highest Multi-hop CE-MRR@20 among the internal configurations, but both full unions complete ten of fifteen questions by rank twenty compared with nine for KG. These descriptive findings motivate choosing the retrieval configuration according to the evidence objective and expected question mixture.

\section{Related Work}
Retrieval-augmented generation separates access to external evidence from the parameters of a language model \citep{rag}. Our task lies at the evidence-retrieval stage: the output is an ordered list of original corpus chunks. We do not evaluate generated answers, answer faithfulness, or end-to-end conversational quality. This distinction matters because an improvement in evidence recall does not by itself demonstrate an improvement in generated responses.

Multi-vector retrieval models, including ColBERT, retain token-level representations and aggregate fine-grained query--document interactions \citep{colbert}. BGE-M3 supports multilingual and multiple retrieval representations \citep{bgem3}. These approaches motivate the use of dense retrieval as a complementary evidence channel. The evaluated SV and EN components use the project's established single-vector and Entity Notes-remapped retrieval procedures; we do not equate those implementations with ColBERT or introduce a new dense architecture.

Graph-augmented retrieval provides explicit access to entities and their connections. GraphRAG develops a graph-based approach to query-focused summarization with community-level information \citep{graphrag}, while LightRAG integrates graph structure into retrieval-augmented generation \citep{lightrag}. Our emphasis is narrower: ontology-conditioned extraction for heterogeneous IKS texts, ambiguity-sensitive entity grounding, and projection back to the original chunk identifiers used by expert annotations. We do not benchmark directly against GraphRAG or LightRAG, and therefore make no empirical superiority claim over those systems.

The language-model components serve different purposes. Qwen3-32B supplies structured extraction \citep{qwen3}; BGE-M3 supplies embedding representations \citep{bgem3}; and BGE-reranker-v2-m3 or Qwen3-Reranker-4B orders candidate passages in the current comparison \citep{bgereranker,qwenembedding}. Reciprocal rank fusion offers a score-scale-independent way to combine ranked evidence lists \citep{rrf}. Our contribution is the domain-grounded integration of these components and its analysis, rather than a new foundation model or rank-fusion rule.

\section{Task and Representation}
Let $\mathcal{C}=\{c_1,\ldots,c_N\}$ denote the original corpus chunks and $Y_q\subseteq\mathcal{C}$ the expert-annotated relevant chunks for question $q$. The retrieval task is to return an ordered list $L_q$ of original chunk identifiers. Internal extraction chunks and graph identifiers are not evaluation targets; they must be projected into the identifier space of $\mathcal{C}$.

We represent the graph as $G=(V,E,\tau,\pi)$, where $\tau$ assigns entity and relation types and $\pi$ associates available graph evidence with source chunks. In storage, source documents, extraction chunks, and domain entities form distinct node roles. Structural links connect chunks to sources and entities to extraction chunks; semantic links represent typed relations between entities. Original chunk identifiers provide the bridge from graph-local evidence to the evaluation corpus. Because not every stored relation retains every original mention, $\pi$ is an available provenance mapping rather than a guarantee of exhaustive provenance.

For domain $d$, the ontology is $\mathcal{O}_d=(\mathcal{L}_d,\mathcal{R}_d,\mathcal{T}_d,\mathcal{N}_d)$: entity labels, relation labels, valid type signatures, and negative constraints. Synonyms provide additional normalization rules. Ontologies are selected by book or genre, with shared schemas where appropriate. Thus, book-conditioned extraction does not imply that every book has a unique ontology. The registry includes the families in Table~\ref{tab:ontology}.

\begin{table}[t]\centering
\caption{Ontology inventory used by the extraction pipeline. Counts describe allowed labels, not observed graph populations.}\label{tab:ontology}
\begin{tabular}{lrr}
\toprule
Ontology family & Entity labels & Relation types \\
\midrule
Purana / Epic & 23 & 80 \\
Mahabharata & 31 & 97 \\
Ayurveda & 36 & 65 \\
Bhagavad Gita & 19 & 33 \\
Dharmashastra & 36 & 61 \\
Natyashastra & 80 & 101 \\
Katha Upanishad & 8 & 12 \\
Manasara & 16 & 22 \\
\bottomrule
\end{tabular}

\end{table}

\section{Ontology-Guided Graph Construction}
\subsection{Source Preparation and Extraction Units}
The extraction pipeline consumes structured text records carrying source references. It removes designated supplementary markers, including appended cross-concept and footnote sections, before segmentation. A recursive splitter constructs extraction chunks with a target size of 3,000 characters and an overlap of 250 characters, using paragraph and sentence boundaries where possible. These are character limits, not token limits. Overlap preserves local context but can produce repeated entity mentions and relations, which are subsequently normalized.

Source, chunk, and entity objects receive internal identifiers. Extraction retains source references and chunk associations so that downstream processing can recover evidence. Checkpoints are written at record boundaries, and raw model responses can be retained in JSONL traces. These mechanisms support recovery and inspection; they do not constitute semantic validation of extracted facts.

\subsection{Two-Pass Extraction with Qwen3-32B}
We used Qwen3-32B to extract graph entities and relations, with eight concurrent second-pass requests. Both passes use temperature zero, disable model thinking, and permit up to 8,192 output tokens. The first pass uses top-$p$ 0.85, top-$k$ 10, and repetition penalty 1.15; the relation pass uses top-$p$ 0.6, top-$k$ 1, and repetition penalty 1.0. The optional named-entity recognition stage was disabled.

In the first pass, the model receives the extraction chunk and the selected ontology. It identifies named entities and domain-relevant concepts, assigns a permitted entity label, and generates a short summary for each entity. Naming instructions retain meaningful diacritics, remove honorifics where appropriate, and discourage inconsistent surface forms. A label and summary provide a local semantic description that can later support entity disambiguation.

In the second pass, extraction is performed for each entity using the full extraction chunk, the first-pass entity record, the extracted entity inventory, and the domain ontology. Candidate relation targets are restricted to the supplied entity names, and relation labels follow the ontology's naming conventions. The prompt distinguishes explicit assertions from co-occurrence, preserves semantic direction, permits an empty relation list, and requests supporting text. Domain guardrails prohibit unsupported kinship, causal, therapeutic, legal, or ritual relations where applicable. They also discourage conflating intended actions with completed actions and collapsing mediated interactions into unsupported binary claims. Concurrency changes throughput rather than the logical inputs to each entity-level request.

The response can include textual details and auxiliary annotations, but not all response fields are persisted by graph upload. Evidence strings are model-selected spans and were not independently verified as quotations. Although the relation pass receives the source chunk, graph extraction can still omit qualifications, select an incorrect direction or target, or generate an unsupported edge.

The ontology is enforced through instructions and downstream processing, not through a formally constrained decoder. JSON handling attempts direct parsing and limited recovery of incomplete responses. Invalid or incomplete entity records may be skipped. Such failures can reduce graph coverage even when the source text contains relevant information.

\subsection{Normalization, Merging, and Graph Storage}
Entity consolidation combines normalized names with semantic similarity. The entity-merging procedure uses MiniLM-L6-v2 name embeddings and a cosine threshold of 0.88, together with normalized-name agreement or a stricter similarity condition of 0.95. A summary-based sense guard can reject incompatible merges at a threshold of 0.55. Representative entities are chosen through the merge procedure, labels are consolidated, and relation synonyms are normalized. These heuristics reduce fragmentation, but neither name similarity nor summary similarity establishes identity with certainty.

Transliteration and alias utilities generate capped sets of spelling variants, while selected epithets supply additional lexical links. An optional cross-book linker creates non-destructive \texttt{SAME\_AS} links using folded names, label agreement, and BGE-based summary similarity. We do not attribute the reported retrieval gains specifically to this optional stage.

Neo4j stores source nodes, extraction chunks, typed entities, and semantic relations. A census of our deployed graph database on 7 November 2026 counted 161,114 nodes and 769,239 directed relationships in total, including structural and provenance links. These deployment totals include structural links and are distinct from the relationship-vector index population reported below; they are not a census of the historical evaluation snapshot. Structural relations include \texttt{PART\_OF} and \texttt{EXTRACTED\_FROM}. The upload path preserves selected evidence fields and chunk references. However, merging relations by endpoint and relation type can overwrite properties for repeated mentions. Consequently, graph provenance is useful for candidate recovery but is not an immutable ledger of every extraction event.

\subsection{Validation and Version Scope}
The evaluated graph was constructed from merged extraction outputs without the optional relation-validation stage. It is therefore not a fully validated or expert-verified graph. Expert curation applies to the retrieval ground truth, not automatically to graph triples. Model-generated relations and their source associations may contain errors, as discussed in the limitations.

The construction records identify Qwen3-32B extraction, bookwise merging, graph upload, and BGE-M3 embedding generation. We distinguish these construction stages from the evaluation of ranked source passages.

\begin{figure*}[t]\centering
\resizebox{0.92\textwidth}{!}{\begin{tikzpicture}[font=\small,box/.style={draw,rounded corners=2pt,align=center,text width=39mm,minimum height=14mm},arrow/.style={-{Stealth[length=2mm]},thick}]
\node[box,fill=KGGreen!8,draw=KGGreen] (text) {Source records\\Genre-specific ontology};
\node[box,fill=KGGreen!8,draw=KGGreen,right=5mm of text] (extract) {Qwen3-32B\\Entities, then relations};
\node[box,fill=KGGreen!8,draw=KGGreen,right=5mm of extract] (graph) {Merge and upload\\Source-linked graph};
\node[box,fill=KGGreen!8,draw=KGGreen,below=15mm of text] (query) {Question interpretation\\Sense-aware grounding};
\node[box,fill=KGGreen!8,draw=KGGreen,right=5mm of query] (retrieve) {Edges, bounded paths\\Lexical and vector search};
\node[box,fill=KGGreen!8,draw=KGGreen,right=5mm of retrieve] (project) {Rank fusion\\Original-chunk projection};
\node[box,fill=KGGreen!8,draw=KGGreen,below=12mm of project] (rerank) {Passage reranking\\Ordered evidence chunks};
\node[box,fill=ENBlue!8,draw=ENBlue,dashed,below=12mm of retrieve] (mv) {Optional SV / mapped EN\\candidate union};
\draw[arrow] (text)--(extract);
\draw[arrow] (extract)--(graph);
\draw[arrow] (query)--(retrieve);
\draw[arrow] (retrieve)--(project);
\draw[arrow] (graph)--(project);
\draw[arrow] (project)--(rerank);
\draw[arrow,dashed] (mv)--(rerank);
\end{tikzpicture}}
\caption{Construction and retrieval pipeline used by the recorded Qwen3-32B generation and KG retrieval runs. Ontologies condition extraction; graph-derived evidence is projected to original corpus chunks before reranking. External SV and mapped EN candidates are combined with KG candidates only in the union configurations; KG-alone uses only graph-derived candidates.}\label{fig:pipeline}
\Description{A directed flow diagram links source records and genre ontologies to Qwen extraction and graph upload, then question grounding, graph search, projection, and passage reranking. Green identifies the graph pipeline; the optional SV / mapped EN candidate union used in union configurations is blue.}
\end{figure*}
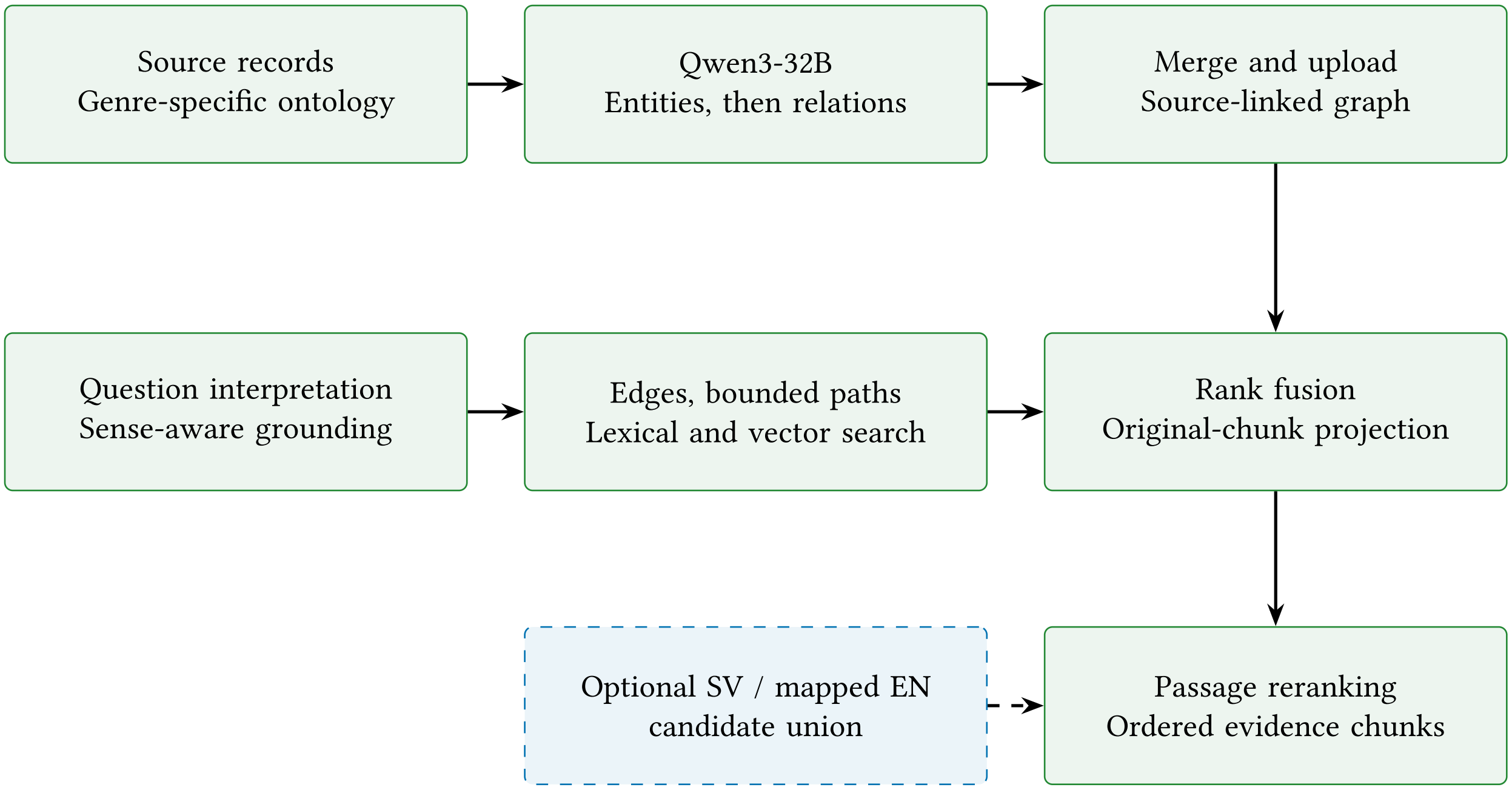

\section{Robust KG Evidence Retrieval}
\subsection{Query Interpretation and Entity Grounding}
The retrieval procedure first interprets a question into entity mentions, canonical forms, possible epithets, sense hints, relation hints, answer type, and textual scope. Structured parsing limits the number of primary entity mentions to three, with heuristic fallback when parsing fails. This representation allows the resolver to combine lexical identity with contextual evidence instead of selecting an entity by name alone.

Candidate entities are collected from exact names and aliases, transliteration variants, full-text lookup, and an entity-vector index. The vector channel retrieves up to eight candidates above a cosine threshold of 0.4; alias and equivalence links can expand the pool. The combined resolver pool is capped at twenty entities. Exact matching is valuable for distinctive names, while semantic and alias channels help when the question uses an alternate form or a descriptive mention.

Candidates are scored using name agreement, type compatibility, summary similarity, textual scope, and relation-neighborhood compatibility, with penalties for mixed senses and known homograph confusions. Schematically,
\begin{equation}
\begin{aligned}
 S(v,q)={}&S_{\mathrm{name}}+S_{\mathrm{type}}\\
 &+0.30\max(0,\cos(\mathbf{s}_v,\mathbf{q}))\\
 &+S_{\mathrm{scope}}+S_{\mathrm{rel}}+S_{\mathrm{exact}}-P(v,q).
\end{aligned}
\end{equation}
The implementation assigns 0.35 to exact folded-name agreement, up to 0.18 to label compatibility, 0.10 to scope agreement, 0.18 to relation compatibility, and 0.08 to an exact-lookup bonus. Penalties and partial matches depend on the case. These are hand-designed ranking signals, not calibrated probabilities. Typically three candidates are retained. Ambiguous cases can expose up to five near-tied candidates to a language-model selection step, which returns a bounded subset. Domain-specific homograph rules supplement this process and are included in the limitations of the evaluation.

\subsection{Local Edges and Bounded Paths}
For grounded entities, retrieval collects either relation-compatible edges or non-structural neighboring edges. Candidate triples and their available evidence details are compared semantically with the interpreted question. The local procedure uses bounded edge counts and retains a small fallback set when similarity filtering would remove all evidence.

When multiple entity groups are present, the robust path channel searches for direct connections and paths of exactly two hops between bounded candidate sets. Source and chunk nodes are excluded as semantic intermediate nodes. The channel caps returned paths and evidence edges; it is not an unrestricted graph-reasoning engine. Such traversal can expose passages linked by a shared entity even when no single passage has a strong dense match to the whole question. Nevertheless, connecting two graph entities does not establish that the resulting path answers the question.

Direction is represented in extracted relations and used by path processing, but not every retrieval channel applies the parsed direction constraint consistently. The reranker therefore remains important for distinguishing an evidential passage from a merely related neighborhood result. We do not interpret every retrieved edge as a logically entailed answer.

\subsection{Lexical Anchors and Relationship-Vector Search}
The robust procedure also collects source chunks associated with selected entities and full-text matches for focused query terms. These lexical anchors are useful when graph extraction has captured an entity but omitted the relation needed by the question. Additional keyword rules identify list-oriented and relation-oriented evidence, with capped full-text candidate lists. These rules use question text, not the ground-truth category label, at retrieval time.

Relationship-vector search provides a second recovery route when entity grounding or local traversal is incomplete. Stored relationship embeddings use BGE-M3 with 1,024 dimensions. The cached relationship matrix contains 319,463 relationship vectors; this is an index population, not a count of all edges in the database. Search computes query--matrix similarities directly rather than using an approximate nearest-neighbor index. It retrieves the top fifty relationship rows and maps them to at most twenty-four distinct original chunk identifiers.

\subsection{Fusion and Evidence Projection}
Ranked graph-evidence lists are combined by reciprocal rank fusion \citep{rrf}:
\begin{equation}
 \operatorname{RRF}(e)=\sum_{j:e\in L_j}\frac{1}{20+\operatorname{rank}_{L_j}(e)}.
\end{equation}
Evidence is keyed by source entity, relation, and target identity or name. The fused head is capped at forty-eight rows, after which relationship-vector candidates can add up to twenty-four rows. These are intermediate evidence limits, not a fixed number of original passages. A row may map to several original chunk identifiers, and duplicate identifiers are removed during projection.

Graph-local chunk identifiers are translated into original corpus identifiers. When direct relation provenance is unavailable, selected entity provenance provides a bounded fallback. Candidates from the KG retrieval procedures are combined before passage reranking. In the KG-only configuration, no separate SV or EN candidates are added; lexical and relationship-vector operations are internal to the KG retriever. This architecture separates graph discovery from the final evidence unit: the last-stage scorer sees original chunk text rather than only a triple serialization.

The evaluated KG retriever is edge-first: it retrieves grounded neighborhood evidence and uses relationship-vector fallback when required, together with the bounded path procedure described above.

\subsection{Reranking and Evidence-Graph Unions}
We rerank candidate chunks with BGE-reranker-v2-m3 and Qwen3-Reranker-4B. For a candidate pool $A_q$, the output is
\begin{equation}
 L_q=\operatorname{sort}_{c\in A_q}\,f_\theta(q,c).
\end{equation}
Candidates are sorted in descending relevance-score order. KG alone reranks only passages returned by the KG retriever; SV alone reranks passages from single-vector retrieval. The full KG + SV and KG + SV + EN configurations select a pool of $N_q$ unique original passages from the respective source pools, where $N_q$ is the KG candidate count for question $q$. EN candidates are mapped to original source chunks before combination. Both full unions rerank the entire selected pool. The pinned variant uses the same selected KG + SV pool but places shared passages first in KG-reranked order and reranks the remaining passages. The selection policies are specified in Section~\ref{sec:results-fixed-budget}. The Qwen path uses the original question, an 8,192-token model limit, and a 1,024-token document cap. Equal candidate counts control passage count, not token volume or total retrieval and construction cost.

\section{Experimental Design}
\subsection{Corpus, Ground Truth, and Coverage}
We use \emph{\dharabenchtitle}~\citep{dharabench}, containing 325 questions with nonempty chunk-level ground truth across twelve categories. The benchmark spans epics, Puranic literature, Ayurveda, philosophy, architecture, performance theory, and historical works. The archived SV/EN and Gemini outputs cover all 325 questions. The KG evaluation outputs cover 320 matched questions; five absent predictions are excluded rather than imputed. The reported comparisons therefore use the exact 320-question intersection. The focal T03, T06, and T07 categories have complete coverage in every system: 22 counting, 19 composite/multi-turn, and 15 multi-hop questions. Table~\ref{tab:categories} reports the full benchmark distribution.

\begin{table*}[t]\centering\small
\caption{Question categories in \emph{\dharabenchtitle}.}\label{tab:categories}
\begin{tabular}{llr}
\toprule
Code & Question category & $n$ \\
\midrule
T01 & Fact Retrieval & 51 \\
T02 & List / Set Retrieval & 40 \\
T03 & Counting / Aggregation & 22 \\
T04 & Chronology \& Sequence understanding & 25 \\
T05 & Lexical Understanding & 31 \\
T06 & Composite / multi-turn & 19 \\
T07 & Multi-hop Reasoning & 15 \\
T08 & Verse \& Textual Retrieval & 26 \\
T09 & False Premise & 18 \\
T10 & Summarization / Expansion / Composition & 29 \\
T11 & Analytical / Interpretive Reasoning & 32 \\
T12 & Mathematical / Logical / Computational Problems & 17 \\
\midrule
Total & & 325 \\
\bottomrule
\end{tabular}

\end{table*}

The number of annotated relevant chunks varies substantially across questions. Thus, a ranking can have high early precision but modest recall when the reference set is large. Precision and recall describe different aspects of the same ranking and should not be substituted for one another. Category labels are used for analysis, not as oracle routing inputs.

\subsection{Metrics and Recalculation}
For all systems, including Gemini, we normalize chunk identifiers and remove duplicate chunk IDs before metric calculation, retaining the first occurrence and the order of the remaining results. We match systems by normalized question text. Let $r_q$ be the rank of the first relevant chunk. For cutoff $k$, reciprocal rank is zero when no relevant chunk is retrieved or the first relevant chunk occurs after $k$; we set $r_q=\infty$ for no hit and use $1/\infty=0$. For $Q$ evaluated questions,
\begin{align}
 \operatorname{MRR}@k&=\frac{1}{|Q|}\sum_{q\in Q}\frac{\mathbb{1}[r_q\leq k]}{r_q},\\
 P@k(q)&=\frac{|L_q[:k]\cap Y_q|}{k},\\
 R@k(q)&=\frac{|L_q[:k]\cap Y_q|}{|Y_q|}.
\end{align}
Precision retains denominator $k$ even for a shorter output; missing positions contribute zero gain to nDCG. All reported aggregates are macro-averages over questions. With binary relevance,
\begin{equation}
 \operatorname{nDCG}@k(q)=
 \frac{\sum_{i=1}^{k}\mathbb{1}[L_{q,i}\in Y_q]/\log_2(i+1)}
 {\sum_{i=1}^{\min(k,|Y_q|)}1/\log_2(i+1)}.
\end{equation}
We calculate these metrics from the saved rankings. The ordinary-relevance figures report MRR@20, Recall@20, Precision@5, and nDCG@10 on the exact 320-question intersection. Complete-evidence scores use the same frozen evidence groups and are defined in Section~\ref{sec:results-complete-evidence}. All reported comparisons are descriptive; we make no statistical-significance claims.

\subsection{Configurations and Interpretation}
The fixed-budget evaluation compares KG alone, SV alone, the fully reranked KG + SV union, and KG + SV + EN under both BGE-reranker-v2-m3 and Qwen3-Reranker-4B. A pinned-intersection KG + SV variant changes ordering on the same selected pool. Each internal system retains the original per-question KG candidate count; the selection policies are specified at the start of Results. \emph{Gemini 3.7 Flash} supplies an archived external reference on the same questions with an unmatched output depth and retrieval budget. Equal candidate counts do not equalize token work, construction cost, or total compute, and do not isolate a causal graph-representation effect.

\subsection{Gemini Generation and Source Mapping}
We use the Gemini generation and source-mapping procedure reported in Section 5.4 of \emph{\dharatitle}~\citep{dhara} (\url{https://arxiv.org/abs/2611.11197}). The \texttt{gemini-3.7-flash} model receives a fixed, cached prompt listing 20 prose-source entries and 13 verse-source entries and requesting up to twenty ranked results in JSON, with thinking level HIGH. Generated passages and references are mapped to original corpus chunks using locally hosted \texttt{google/gemma-4-31B-it}, served through vLLM at temperature zero. Reference matching first identifies up to ten table-of-contents locations, from which Gemma selects up to three. BM25 and all-MiniLM-L6-v2 similarity then retrieve candidate chunks, and Gemma selects a matching chunk from up to twenty candidates.

The source mappings were manually reviewed with the question, generated passage and reference, automatic mapping, candidate chunks, and benchmark relevance annotations visible. Reviewers could select a source chunk outside the automatic candidate list or mark a result irrelevant; recorded manual decisions replaced automatic mappings. Consequently, this reference evaluates generation plus mapping and annotation-visible review, not an unassisted Gemini retriever. We reuse these reviewed mappings rather than rerunning generation. For the present comparison, duplicate chunk IDs are removed before scoring, consistently with the internal systems; this differs from the rank-preserving duplicate treatment described in the companion paper. Invalid or unmapped results receive no relevance credit. We restrict scoring to the common 320-question cohort. Gemini's retrieval budget remains unmatched.

\section{Results}

\label{sec:results-fixed-budget}
This section reports the fixed-candidate-budget evaluation on the same 320 questions and frozen Dhara-Bench v3 ground truth. The primary comparisons distinguish KG alone, single vector (SV) alone, the KG + SV union, and the KG + SV + Entity Notes (EN) union. Each question uses its original KG budget $N_q$ in every primary system, giving 27,142 candidate pairs per reranker and a mean budget of 84.81875. EN candidates are the original passage IDs mapped from retrieved insights. The full KG + SV pool contains all shared IDs followed by balanced top exclusive candidates; the EN union prioritizes three-way intersections, then exactly two-way intersections, then single-source IDs, cycling through KG, SV, and EN within each tier until $N_q$ unique passages have been selected. Both full unions rerank the entire selected pool. The pinned KG + SV variant uses the identical selected pool but fixes its shared IDs first in KG reranked order and reranks only the remainder. All reported scores are macro averages over questions. The ordinary-relevance figures report MRR@20, Recall@20, Precision@5, and nDCG@10; Precision@5 is computed directly from saved rankings and retains denominator five. All fixed-budget comparisons are descriptive.

\subsection{KG concentrates early relevance; unions broaden coverage}
\begin{figure*}[t]\centering
\includegraphics[width=\linewidth]{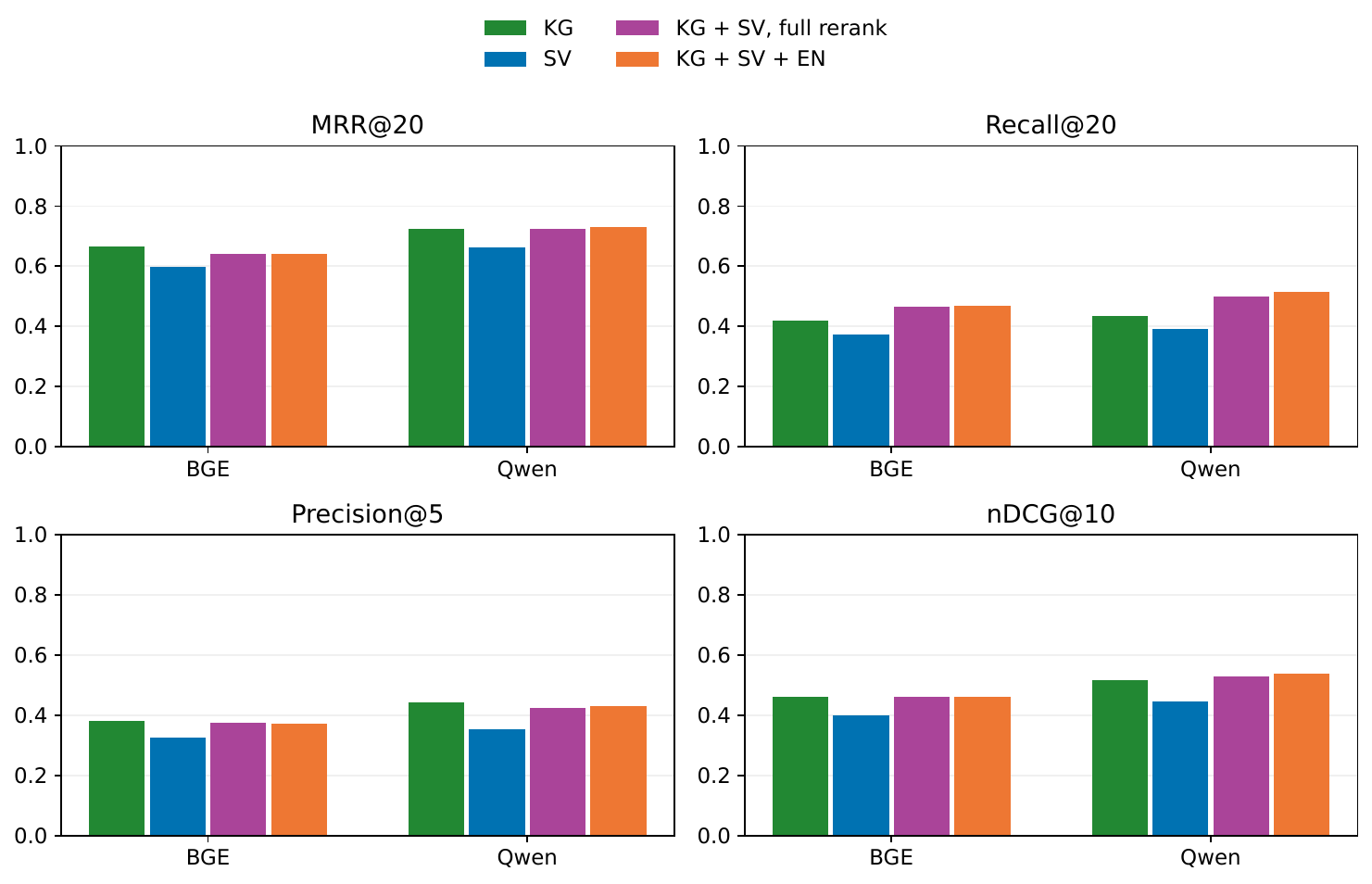}
\caption{Overall scores on the same 320 questions. Each internal system has exactly the original KG candidate budget. Bars compare KG, SV, the fully reranked KG + SV union and the KG + SV + EN union under BGE and Qwen.}\label{fig:fixed-01-overall}
\Description{Overall scores on the same 320 questions. Each internal system has exactly the original KG candidate budget. Bars compare KG, SV, the fully reranked KG + SV union and the KG + SV + EN union under BGE and Qwen.}
\end{figure*}

Figure~\ref{fig:fixed-01-overall} compares the overall quality of the four primary internal systems.

\begingroup\emergencystretch=2em\relax
KG is a strong standalone retriever, but the preferred architecture depends on the metric. With BGE, KG has the highest overall MRR@20 (0.6665) and Precision@5 (0.3806) among the four main internal systems. Adding SV and EN raises Recall@20 from 0.4190 to 0.4693, while MRR@20 falls to 0.6413 and Precision@5 to 0.3731. The union therefore finds a broader set of annotated evidence without preserving the same concentration of relevance at the earliest ranks.\par\endgroup

\begingroup\emergencystretch=2em\relax
The Qwen results show that the standalone KG advantage is metric-dependent. KG retains the highest Precision@5 (0.4419), whereas KG + SV + EN has the highest overall MRR@20 (0.7306), Recall@20 (0.5134) and nDCG@10 (0.5376). Its MRR advantage over KG (0.7253) is small and descriptive. Across the 12 categories, KG beats the full KG + SV union on MRR@20 in eight categories with BGE and seven with Qwen; against the EN union, those counts are eight and five. Thus, KG generally concentrates early relevance, while the category-level winner depends on the reranker and metric.\par\endgroup

\subsection{KG strength on Counting, Composite and Multi-hop questions}
\begin{figure*}[t]\centering
\includegraphics[width=\linewidth]{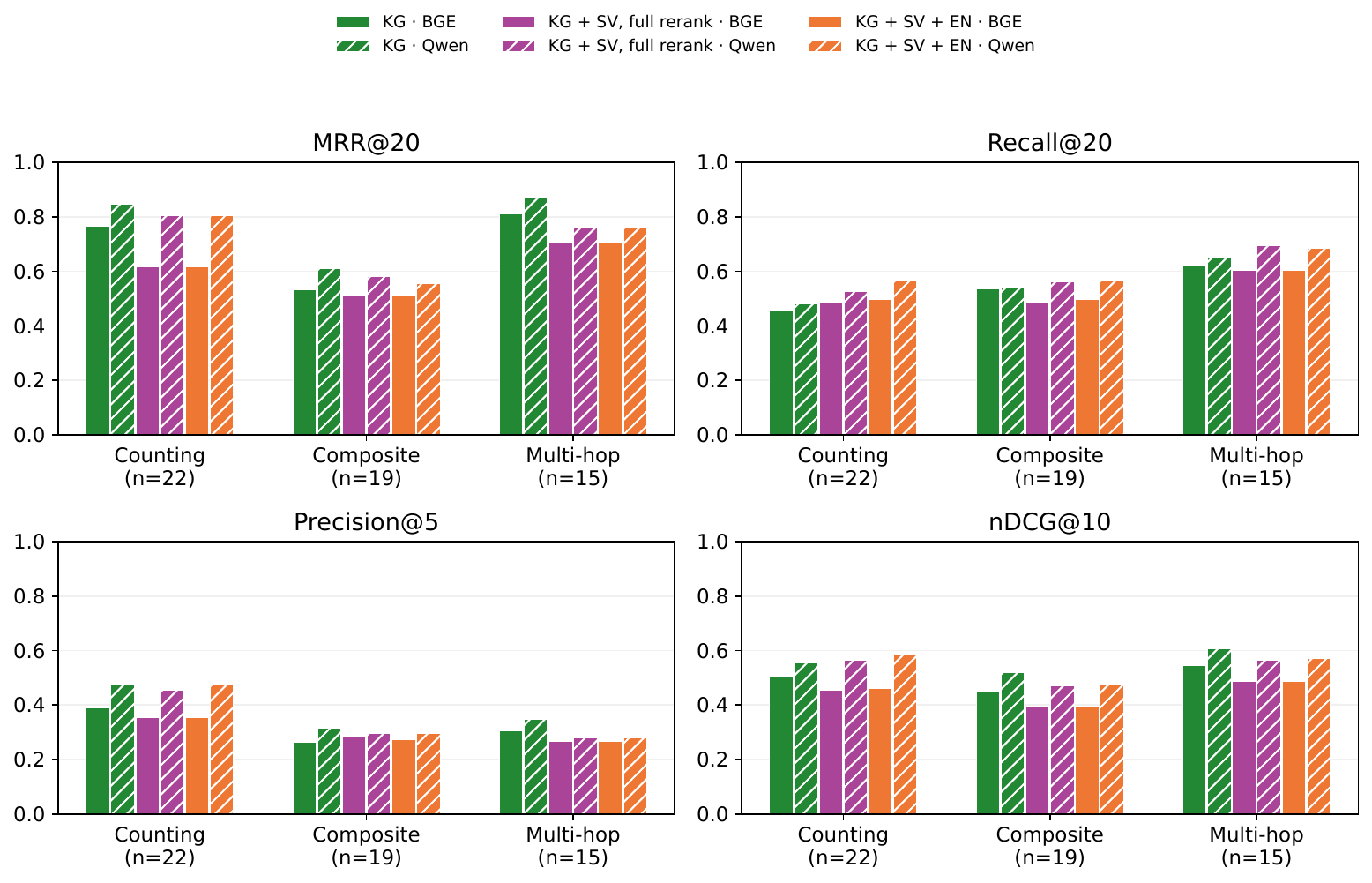}
\caption{Counting (n=22), Composite (n=19) and Multi-hop (n=15), with the same questions and budgets for all systems. Colors identify systems; hatched bars identify Qwen. MRR dominance by KG coexists with recall and selected Precision@5/nDCG exceptions.}\label{fig:fixed-03-focal}
\Description{Counting (n=22), Composite (n=19) and Multi-hop (n=15), with the same questions and budgets for all systems. Colors identify systems; hatched bars identify Qwen. MRR dominance by KG coexists with recall and selected Precision@5/nDCG exceptions.}
\end{figure*}

Figure~\ref{fig:fixed-03-focal} compares the three focal question categories under both rerankers.

\begingroup\emergencystretch=2em\relax
The focal categories show the clearest standalone KG advantage in first-hit ranking. KG has higher MRR@20 than SV, either KG + SV merging policy, and the EN union in Counting, Composite and Multi-hop under both rerankers. With Qwen, KG MRR@20 is 0.8450 for Counting, 0.6111 for Composite and 0.8722 for Multi-hop; the corresponding full-union values are 0.8046, 0.5826 and 0.7622. KG also has the highest nDCG@10 in Composite and Multi-hop among these internal configurations under both models.\par\endgroup

\begingroup\emergencystretch=2em\relax
The KG advantage in these categories is specific to the ranking objective. Qwen Counting nDCG@10 improves from 0.5554 with KG to 0.5857 with the EN union, and Counting Precision@5 ties between KG and the EN union at 0.4727. Qwen Recall@20 is higher with a union in all three focal categories. BGE Composite Precision@5 is higher with the full union (0.2842) than KG (0.2632). The structural organization of KG evidence is consistent with its strong first-hit ranking on relational and aggregation questions, but this experiment does not separately isolate graph traversal, extraction, entity normalization, or candidate generation as causal contributors.\par\endgroup

\subsection{Complete-Evidence Retrieval for Composite and Multi-hop Questions}
\label{sec:results-complete-evidence}
Ordinary relevance metrics need not establish that all evidence required by a question has been recovered. We therefore evaluate the frozen Dhara Bench evidence groups~\citep{dharabench} on the same saved rankings. All 19 Composite/Multi-turn and 15 Multi-hop questions have at least one annotated group, giving 401 alternative groups, including 35 singleton routes. Members within a group are jointly required; different groups provide alternative sufficient evidence sets. A group's completion rank is the latest first-occurrence rank among its members, and the question's completion rank is the earliest completion across its groups. CE-Recall@$k$ is one when a complete group occurs within the first $k$ passages and zero otherwise. CE-MRR@20 averages the reciprocal completion rank, assigning zero if no group is completed within twenty. Scores are macro-averaged over questions, with denominators 19 and 15 rather than all 320 questions or all 401 groups. For the nonempty collection of annotated evidence groups $\mathcal{G}_q$, let $\rho_q(c)$ be the first rank of chunk $c$, or $\infty$ if it is absent. Then
\begin{align}
 t_q &= \min_{g\in\mathcal{G}_q}\max_{c\in g}\rho_q(c),\\
 \operatorname{CE\!\text{-}\!Recall}@k &= \frac{1}{|Q_G|}\sum_{q\in Q_G}\mathbb{1}[t_q\leq k],\\
 \operatorname{CE\!\text{-}\!MRR}@k &= \frac{1}{|Q_G|}\sum_{q\in Q_G}\frac{\mathbb{1}[t_q\leq k]}{t_q},
\end{align}
where $Q_G$ is the evaluated group-annotated category and $1/\infty=0$. These metrics follow \dhara{} Bench~\citep{dharabench} and the companion retrieval study~\citep{dhara}.

\begin{figure*}[t]\centering
\includegraphics[width=\linewidth]{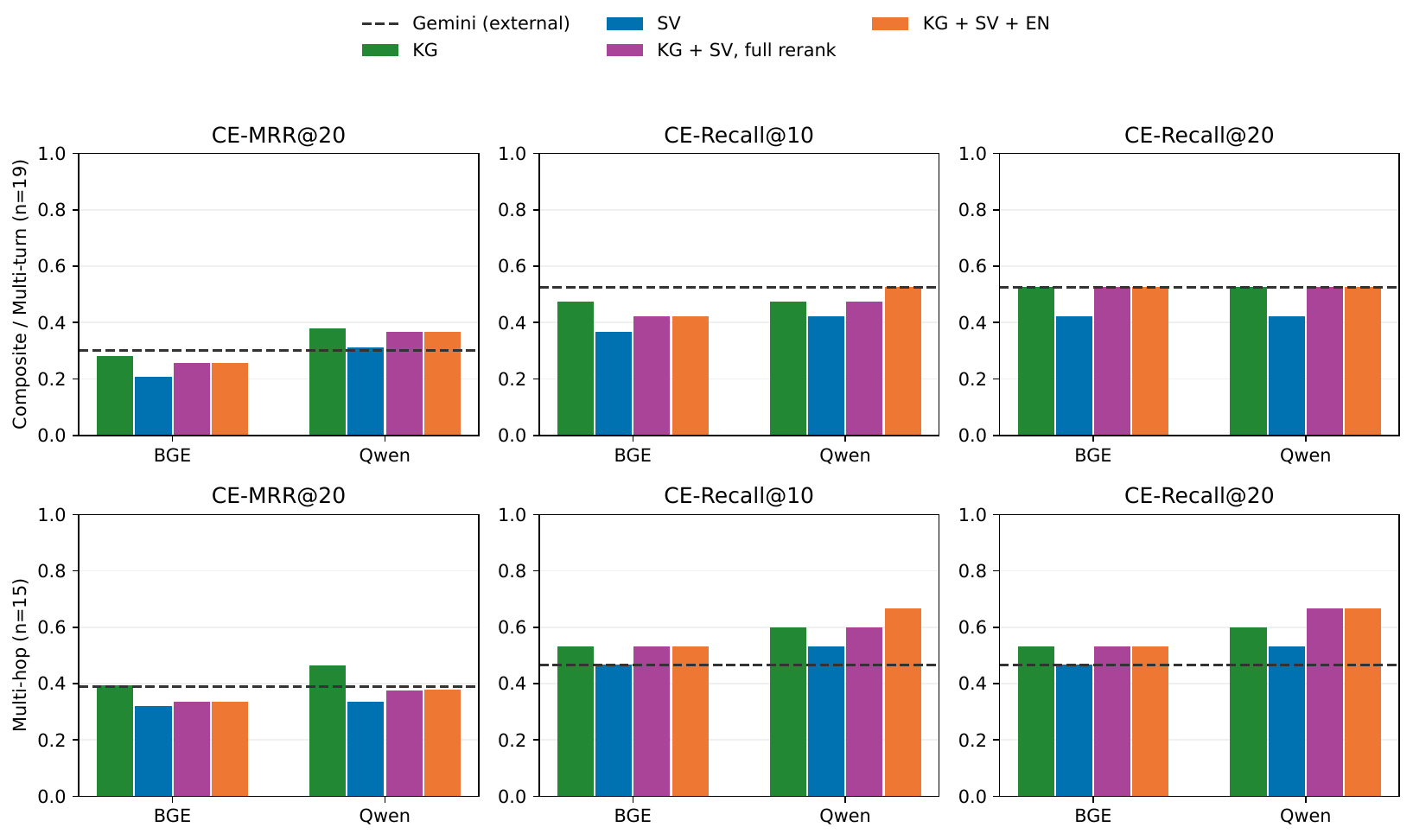}
\caption{Complete-evidence ranking and recovery for Composite/Multi-turn (top row) and Multi-hop (bottom row). Colors identify the four primary fixed-budget internal systems; BGE and Qwen are grouped on the horizontal axes. Dashed lines show the same archived Gemini reference in both model groups. KG's earlier average completion coexists with selected Qwen-union gains in the fraction of questions completed.}\label{fig:fixed-complete-evidence}
\Description{Six panels show complete-evidence MRR at twenty and complete-evidence recall at ten and twenty for Composite and Multi-hop questions. KG leads the four primary internal systems on completion MRR under both models, while Qwen unions complete more Multi-hop questions within twenty results.}
\end{figure*}

Table~\ref{tab:fixed-complete-evidence} and Figure~\ref{fig:fixed-complete-evidence} show a consistent standalone KG advantage over SV in completion ranking. Composite CE-MRR@20 increases from 0.2076 to 0.2803 with BGE and from 0.3114 to 0.3787 with Qwen. The corresponding Multi-hop increases are 0.3194 to 0.3944 and 0.3372 to 0.4651. Among the four primary internal systems, KG has the highest CE-MRR@20 in both categories under both rerankers. At twenty results, KG completes 10/19 Composite questions with either model, compared with 8/19 for SV; Multi-hop completion is 8/15 versus 7/15 with BGE and 9/15 versus 8/15 with Qwen.

Completion speed and the number of completed questions nevertheless favor different configurations. With Qwen, both full unions complete 10/15 Multi-hop questions by rank twenty, exceeding KG's 9/15, but their CE-MRR@20 values (0.3757 for KG + SV and 0.3787 with EN) are lower than KG's 0.4651. The EN union completes all ten of those questions by rank ten, whereas KG and the KG + SV union complete nine by that cutoff. For Composite questions, the Qwen EN union improves CE-Recall@10 from KG's 9/19 to 10/19, while both reach 10/19 at twenty and KG retains higher CE-MRR. Thus, broader coverage can complete more questions within a fixed result window without delivering the earliest sufficient evidence on average.

Against the external reference, Qwen KG has higher CE-MRR@20 in both Composite (0.3787 versus 0.3026) and Multi-hop (0.4651 versus 0.3889). Gemini completes 10/19 Composite questions by rank ten, matching the Qwen EN union and exceeding KG's 9/19; at twenty, all three tie. For Multi-hop, the Qwen EN union completes 10/15 questions at both cutoffs, compared with Gemini's 7/15. These are descriptive evidence-retrieval comparisons: Gemini's depth and candidate budget are not matched, and neither complete-set retrieval nor singleton completion establishes answer correctness or a particular reasoning process.

\begin{table*}[t]\centering\small
\caption{Complete-evidence results on the same 19 Composite/Multi-turn and 15 Multi-hop questions. All internal configurations use the exact per-question KG candidate budget; Gemini is an archived external reference with an unmatched retrieval budget. CE-MRR is truncated at twenty, and CE-R is the fraction of questions with at least one complete annotated evidence group at the indicated cutoff.}\label{tab:fixed-complete-evidence}
\begin{tabular}{llrrrrrr}\toprule
& & \multicolumn{3}{c}{Composite / Multi-turn ($n=19$)} & \multicolumn{3}{c}{Multi-hop ($n=15$)} \\
\cmidrule(lr){3-5}\cmidrule(lr){6-8}
System & Reranker & CE-MRR@20 & CE-R@10 & CE-R@20 & CE-MRR@20 & CE-R@10 & CE-R@20 \\\midrule
KG & BGE & 0.2803 & 0.4737 & 0.5263 & 0.3944 & 0.5333 & 0.5333 \\
SV & BGE & 0.2076 & 0.3684 & 0.4211 & 0.3194 & 0.4667 & 0.4667 \\
KG + SV (full) & BGE & 0.2557 & 0.4211 & 0.5263 & 0.3361 & 0.5333 & 0.5333 \\
KG + SV (pinned) & BGE & 0.2470 & 0.4211 & 0.5263 & 0.4111 & 0.5333 & 0.5333 \\
KG + SV + EN & BGE & 0.2557 & 0.4211 & 0.5263 & 0.3361 & 0.5333 & 0.5333 \\
\midrule
KG & Qwen & 0.3787 & 0.4737 & 0.5263 & 0.4651 & 0.6000 & 0.6000 \\
SV & Qwen & 0.3114 & 0.4211 & 0.4211 & 0.3372 & 0.5333 & 0.5333 \\
KG + SV (full) & Qwen & 0.3666 & 0.4737 & 0.5263 & 0.3757 & 0.6000 & 0.6667 \\
KG + SV (pinned) & Qwen & 0.3338 & 0.4211 & 0.5263 & 0.4556 & 0.5333 & 0.5333 \\
KG + SV + EN & Qwen & 0.3671 & 0.5263 & 0.5263 & 0.3787 & 0.6667 & 0.6667 \\
\midrule
Gemini & --- & 0.3026 & 0.5263 & 0.5263 & 0.3889 & 0.4667 & 0.4667 \\
\bottomrule\end{tabular}
\end{table*}

\subsection{KG Shows Opposite Trends in Precision and Recall Relative to Unions}
\begin{figure*}[t]\centering
\includegraphics[width=\linewidth]{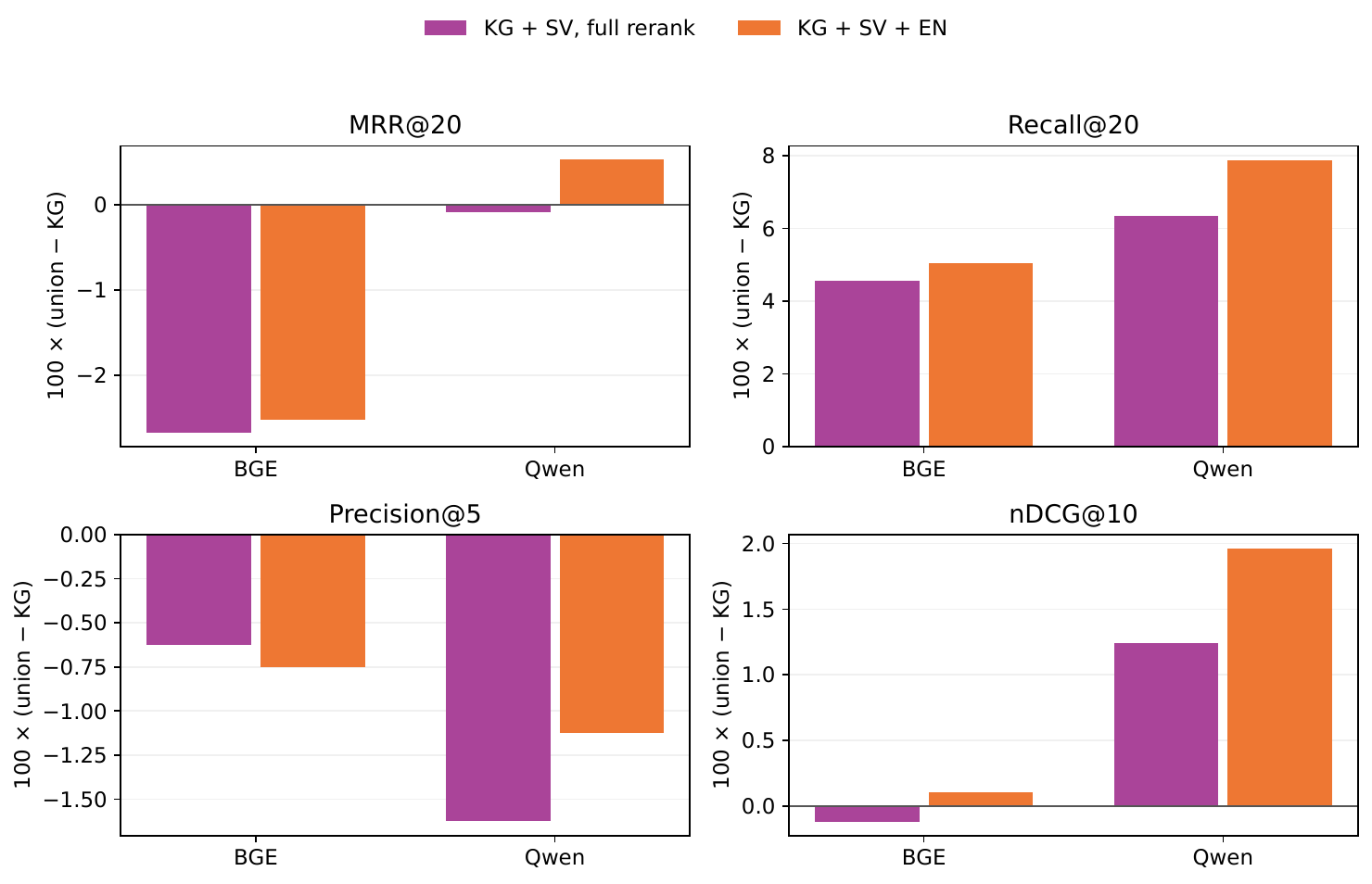}
\caption{Union minus KG differences, multiplied by 100 for readability. Positive values favor a union; negative values favor KG. Differences are score units \ensuremath{\times}100, not relative percentage improvements. These descriptive deltas do not establish statistical significance.}\label{fig:fixed-04-tradeoff}
\Description{Union minus KG differences, multiplied by 100 for readability. Positive values favor a union; negative values favor KG. Differences are score units \ensuremath{\times}100, not relative percentage improvements. These descriptive deltas do not establish statistical significance.}
\end{figure*}

Figure~\ref{fig:fixed-04-tradeoff} shows the precision--recall contrast through union-minus-KG differences.

\begingroup\emergencystretch=2em\relax
KG shows opposite overall trends in early precision and deeper recall relative to the unions: it concentrates relevant passages among the first five results, while the unions recover a broader set of annotated evidence within twenty results. For BGE, the full union gains 0.0456 Recall@20 over KG but loses 0.0268 MRR@20 and 0.0063 Precision@5. For Qwen, the EN union gains 0.0787 Recall@20, while Precision@5 decreases by 0.0113. This observed precision--recall contrast is metric- and category-dependent; Qwen MRR@20 and nDCG@10 also improve with the EN union, so an increase in coverage need not reduce every early-ranking metric.\par\endgroup

\begingroup\emergencystretch=2em\relax
These metrics answer different questions: MRR@20 rewards the earliest relevant hit, Precision@5 measures the proportion of relevant passages among the first five results, Recall@20 measures coverage of the annotated relevant set, and nDCG@10 accounts for the positions of relevant items in the first ten. The results are consistent with KG concentrating annotated relevant chunks near the top and unions expanding coverage. Architecture selection should identify whether the application values the first useful passage, early precision, or broader evidence coverage.\par\endgroup

\subsection{Question Composition and Hardware Should Guide Architecture Choice}
\begin{figure*}[t]\centering
\includegraphics[width=0.94\linewidth]{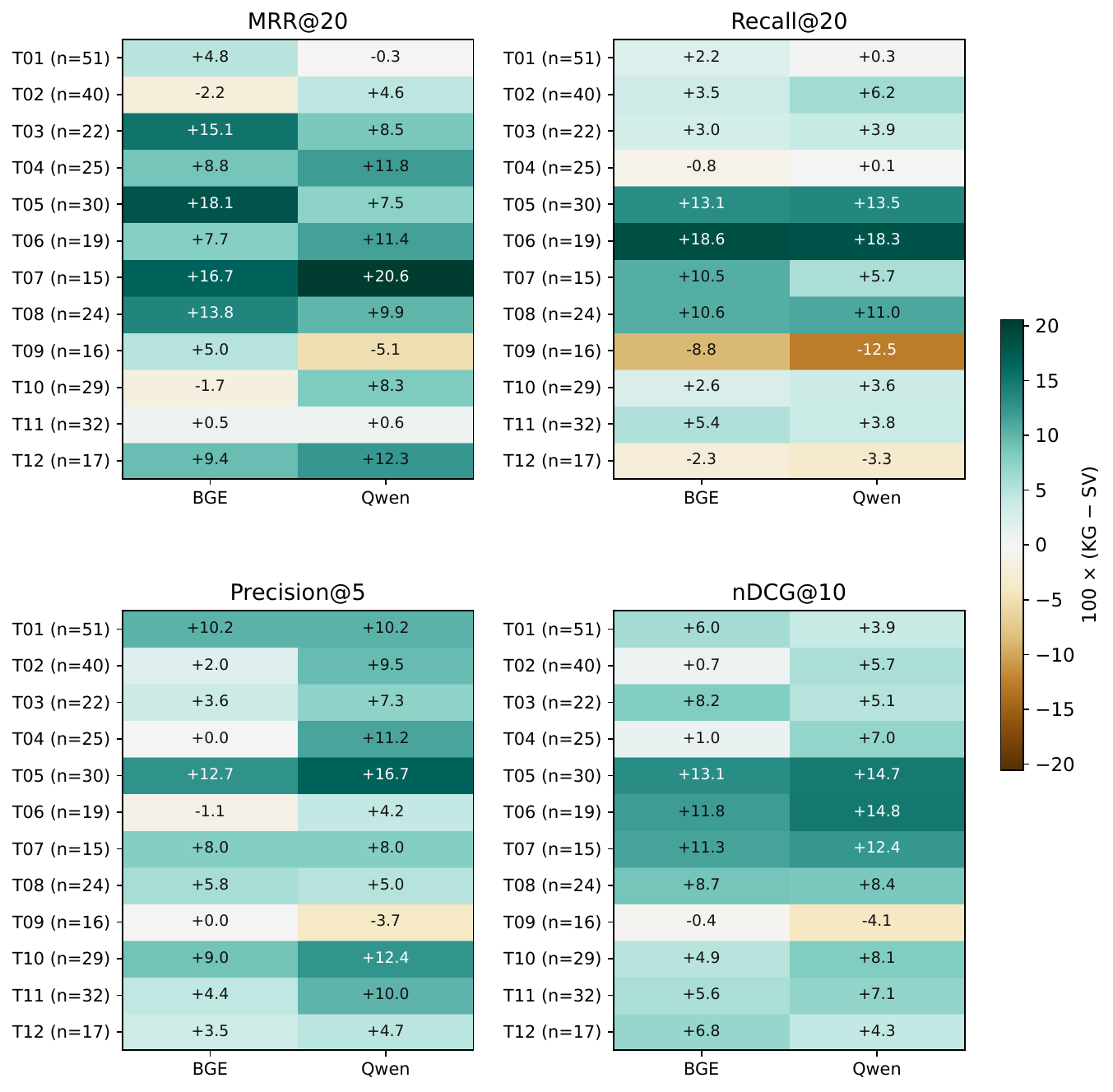}
\caption{Category-level KG minus SV deltas (score \ensuremath{\times}100). Positive cells favor KG; negative cells favor SV. Labels give category codes and sample sizes. Each cell is an observed category mean, not a validated routing policy.}\label{fig:fixed-05-routing}
\Description{Category-level KG minus SV deltas (score \ensuremath{\times}100). Positive cells favor KG; negative cells favor SV. Labels give category codes and sample sizes. Each cell is an observed category mean, not a validated routing policy.}
\end{figure*}

Figure~\ref{fig:fixed-05-routing} shows the category-specific differences between KG and SV.

\begingroup\emergencystretch=2em\relax
KG improves on SV in most category-level comparisons: MRR@20 is higher in 10 of 12 categories with either reranker, and nDCG@10 is higher in 11 of 12. Nevertheless, the category exceptions matter for a heterogeneous question pool. In False Premise (T09; n=16), Qwen SV exceeds KG on all four metrics. With BGE, SV has higher MRR@20 in List\slash Set Retrieval (T02) and Summarization\slash Expansion\slash Composition (T10). With Qwen, SV also has higher Fact Retrieval MRR@20 (T01). Other exceptions concern individual metrics, such as recall in Mathematical\slash Logical\slash Computational Problems (T12).\par\endgroup

\begingroup\emergencystretch=2em\relax
This heterogeneity supports testing category-aware routing: emphasize KG for pools dominated by the focal structural tasks, retain SV as a candidate architecture for categories where it performs well, and evaluate unions when evidence coverage is the primary objective. These are hypotheses for held-out routing experiments; the present study has not trained or evaluated a router. The runtime study in Section~\ref{sec:results-timing} quantifies the retrieval overhead on a matched-question, matched-candidate sample. Routing decisions should be judged on the actual category mixture and quality objective, rather than assuming a single system dominates every workload.\par\endgroup

\begingroup\emergencystretch=2em\relax
Architecture selection should also depend on the target hardware. KG generation with Qwen3-32B requires substantial compute before serving queries, and passage reranking adds accelerator work at query time. Available GPU memory, scoring throughput, concurrent query load, graph storage, and the frequency of corpus updates therefore constrain a practical choice among KG, SV, and their unions. A stable corpus can amortize graph construction over repeated queries, whereas a hardware-constrained service should weigh the measured quality gains against the resources required for construction and reranking. These are deployment implications of the evaluated pipeline; the quality experiment does not establish a hardware-specific latency or cost threshold.\par\endgroup

\subsection{Qwen Generally Improves Quality with a Time Trade-off; BGE Has Local Exceptions}
\begin{figure*}[t]\centering
\includegraphics[width=0.94\linewidth]{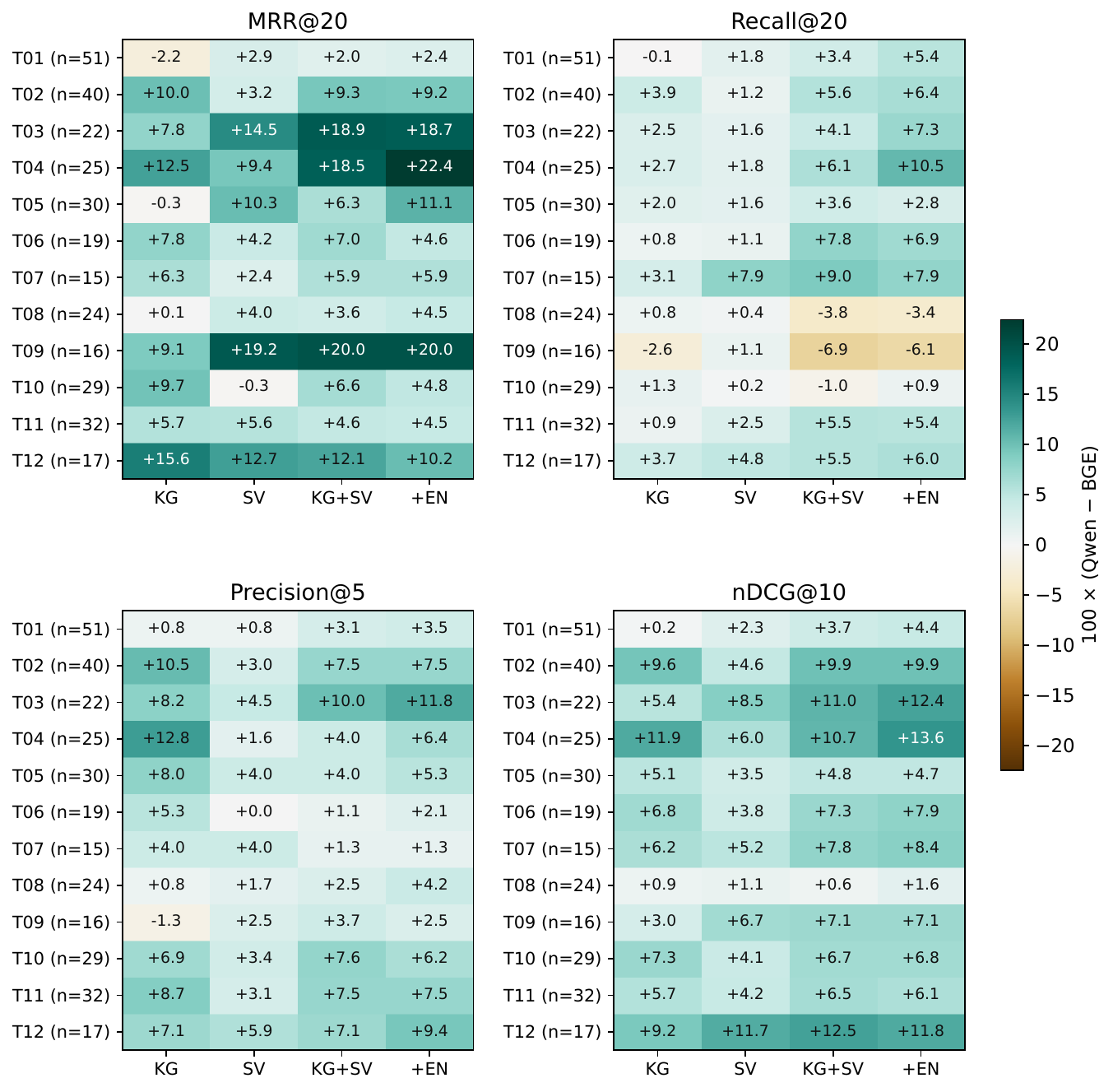}
\caption{Qwen minus BGE differences by category and system (score \ensuremath{\times}100). Positive cells favor Qwen; negative cells favor BGE. KG+SV denotes the exact-budget full-rerank union. +EN denotes KG + SV + mapped EN original passages.}\label{fig:fixed-06-rerankers}
\Description{Qwen minus BGE differences by category and system (score \ensuremath{\times}100). Positive cells favor Qwen; negative cells favor BGE. KG+SV denotes the exact-budget full-rerank union. +EN denotes KG + SV + mapped EN original passages.}
\end{figure*}

Figure~\ref{fig:fixed-06-rerankers} shows where Qwen improves quality and where BGE has local exceptions.

\begingroup\emergencystretch=2em\relax
Across all four main systems, Qwen has higher overall scores than BGE on all four reported metrics. For KG alone, MRR@20 increases from 0.6665 to 0.7253, Precision@5 from 0.3806 to 0.4419, and nDCG@10 from 0.4616 to 0.5180. Qwen also improves the focal-category MRR results and yields the largest overall Recall@20 when combined with KG + SV + EN (0.5134). On this evaluated workload, Qwen is therefore the stronger general quality choice.\par\endgroup

\begingroup\emergencystretch=2em\relax
The category matrix also records cases for retaining BGE in an empirical comparison. Within KG, BGE exceeds Qwen on Fact Retrieval MRR@20 and Recall@20, Lexical Understanding MRR@20, and False Premise Recall@20 and Precision@5. Within the full union, BGE has higher Recall@20 for Verse\slash Textual Retrieval, False Premise and Summarization\slash Expansion\slash Composition. These exceptions make reranker selection dependent on the category and metric. The A100 measurements in Section~\ref{sec:results-timing} show approximately 7.3-fold longer reranking with Qwen than BGE, quantifying the time trade-off accompanying its general quality advantage. BGE remains relevant when its category-specific quality is adequate and the deployment has restrictive time or compute budgets. Concurrent end-to-end serving throughput and operating cost remain outside the present timing study.\par\endgroup

\subsection{Retrieval and Reranking Time}
\label{sec:results-timing}
Runtime was measured on 24 questions selected to cover all twelve categories and ten domains, with two independently shuffled passes (48 paired trials). Each SV request returned exactly the number of passages obtained from the corresponding Robust KG request: 52--141 candidates, with a mean of 82.79. KG query embeddings and the 319,463-vector relationship matrix ran on NVIDIA A100 80 GB GPU 2; SV embeddings ran on GPU 0 of the same A100 host. Retrieval was timed without passage reranking, and the saved pools were subsequently scored with BGE and Qwen on the same A100. All paired requests and scoring runs succeeded. This category-balanced sample characterizes the deployed pipeline rather than the full 320-question runtime distribution.

\begin{figure*}[t]\centering
\includegraphics[width=\linewidth]{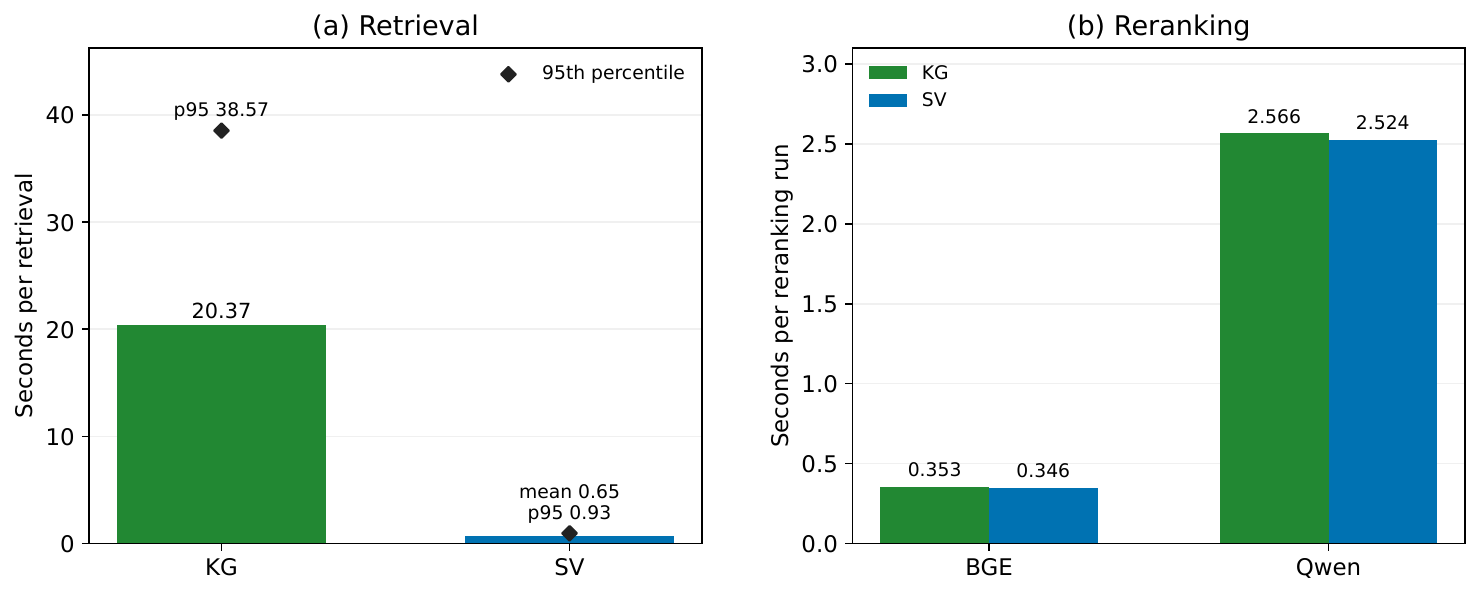}
\caption{GPU-backed timing on 24 questions measured twice at identical per-trial candidate budgets. (a) Mean retrieval time, with diamonds marking the 95th percentile. (b) Mean fresh passage-reranking time on the retrieved KG and SV pools. Loading and explicit reranker warm-up are excluded; the two stages were measured separately.}\label{fig:runtime-stages}
\Description{Two timing panels compare KG and SV. KG retrieval averages 20.374 seconds versus 0.646 for SV. BGE reranking takes about 0.35 seconds for either pool and Qwen about 2.55 seconds. Diamonds mark retrieval 95th percentiles.}
\end{figure*}

Figure~\ref{fig:runtime-stages} separates retrieval from reranking. Mean retrieval time was 20.374 s for KG and 0.646 s for SV, a 31.6-fold difference and an absolute gap of 19.728 s. The respective 95th percentiles were 38.566 s and 0.935 s. KG's first- and second-pass means were 21.004 s and 19.743 s, indicating that model residency did not eliminate the retrieval gap. The observed difference includes query interpretation, entity resolution, graph and relationship search, provenance projection, and passage fetching; it is not an isolated measurement of graph traversal.

BGE (FP16) reranking averaged 0.353 s for KG pools and 0.346 s for SV pools, compared with 2.566 s and 2.524 s for Qwen (BF16), approximately 7.3 times longer. BGE and Qwen used maximum batch sizes of 16 and 8, respectively, with passages capped at 1,024 model-specific tokens; model loading and explicit warm-up were excluded. Reranking time correlated more closely with capped document-token volume (Pearson $r=0.973$--$0.987$) than candidate count ($r=0.796$--$0.903$). Equal candidate budgets therefore control passage count but not exact scoring work. Summing the separately measured stages gives mean KG/SV times of 20.727/0.992 s with BGE and 22.939/3.170 s with Qwen. Retrieval contributes 88.8\% of the KG+Qwen total, whereas reranking contributes 79.6\% of the SV+Qwen total. These composed times exclude answer generation and were measured sequentially on shared hardware, not through an integrated end-to-end serving request.

\subsection{KG Matches or Beats Gemini in Specific Categories and Metrics, with Losses Elsewhere}
\begin{figure*}[t]\centering
\includegraphics[width=\linewidth]{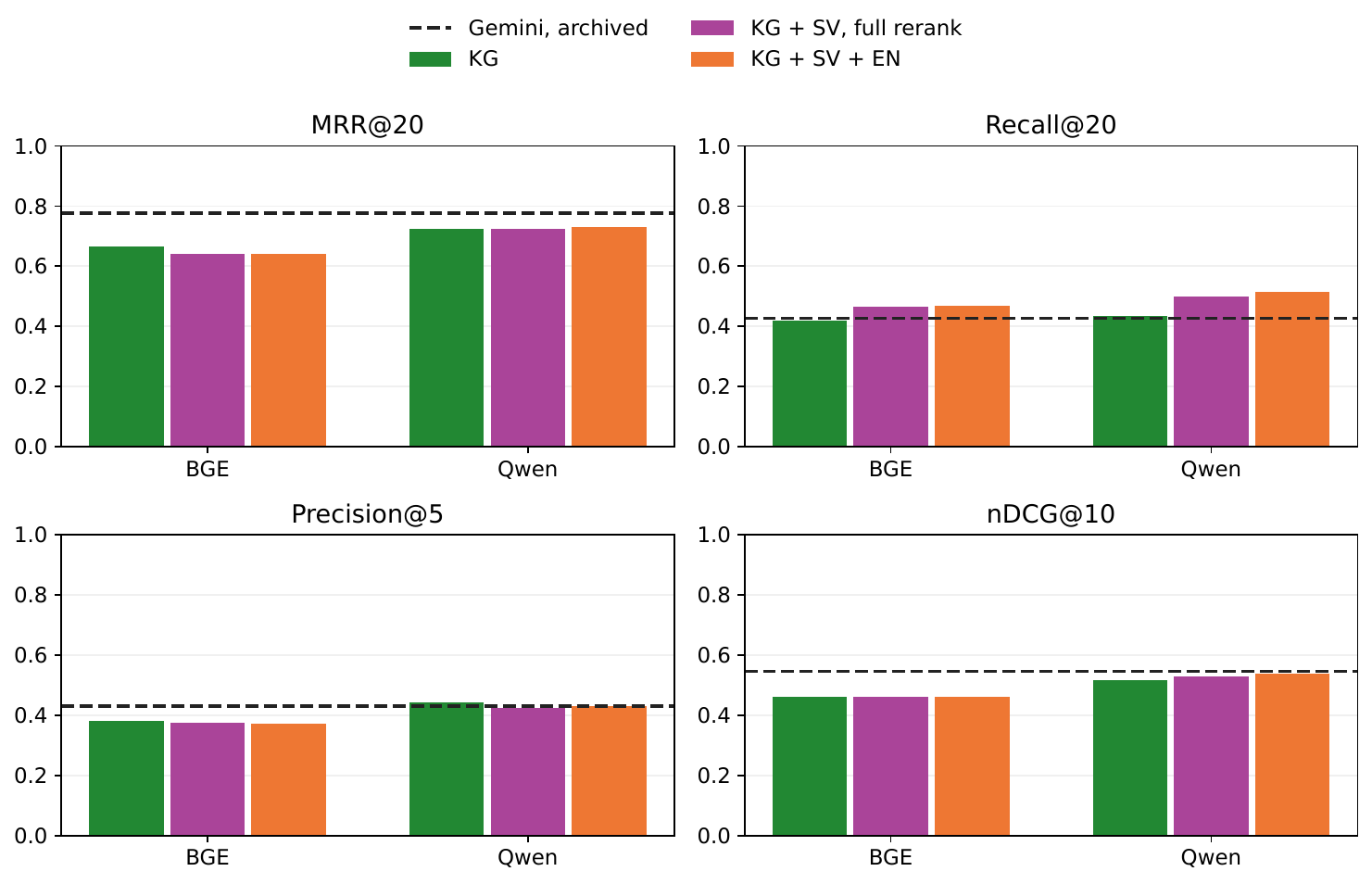}
\caption{Final external comparison on the matched 320-question cohort. Dashed lines show the same archived Gemini reference in both reranker groups. Internal bars use exact KG budgets; the external retrieval budget is not matched. The result is a coverage--ranking comparison rather than an across-the-board winner.}\label{fig:fixed-07-gemini}
\Description{Final external comparison on the matched 320-question cohort. Dashed lines show the same archived Gemini reference in both reranker groups. Internal bars use exact KG budgets; the external retrieval budget is not matched. The result is a coverage--ranking comparison rather than an across-the-board winner.}
\end{figure*}

Figure~\ref{fig:fixed-07-gemini} compares the internal systems with the archived Gemini reference.

\begingroup\emergencystretch=2em\relax
The external comparison uses the reviewed Gemini source mappings described in Experimental Design, restricted to the same 320 questions and ground truth. Duplicate chunk IDs are removed before metric calculation for both Gemini and the internal systems. Invalid or unmapped results receive no relevance credit; shorter rankings are not padded with additional retrieval, and Precision@5 retains denominator five. This is a matched-cohort comparison, but Gemini's output depth and retrieval budget differ from the internal systems, and its mappings include annotation-visible manual review.\par\endgroup

\begingroup\emergencystretch=2em\relax
The internal systems match or exceed Gemini in specific question categories and metrics, with losses elsewhere. Gemini has higher overall MRR@20 (0.7767) and nDCG@10 (0.5459) than every internal configuration. KG + SV + EN with Qwen provides higher Recall@20 (0.5134 versus 0.4272), and its Precision@5 ties Gemini at 0.4306. Qwen KG alone has higher Precision@5 (0.4419) than Gemini. Thus, the strongest internal coverage result does not imply superiority on all ranking metrics.\par\endgroup

\begingroup\emergencystretch=2em\relax
Category-level results sharpen the comparison. In Composite, Qwen KG exceeds Gemini on all four metrics: MRR@20 0.6111 versus 0.5789, Recall@20 0.5428 versus 0.3833, Precision@5 0.3158 versus 0.3053, and nDCG@10 0.5183 versus 0.4459. In Multi-hop, Qwen KG has slightly higher MRR@20 (0.8722 versus 0.8667) and higher Recall@20 (0.6508 versus 0.5517), ties Precision@5 at 0.3467, and has slightly lower nDCG@10 (0.6071 versus 0.6107). Counting favors Gemini on MRR@20, Precision@5 and nDCG@10, while the Qwen EN union provides greater Recall@20 (0.5695 versus 0.4814). These descriptive results establish category- and metric-specific matches, gains, and losses against the external reference, rather than a universal winner.\par\endgroup

\begin{figure*}[t]\centering
\includegraphics[width=\linewidth]{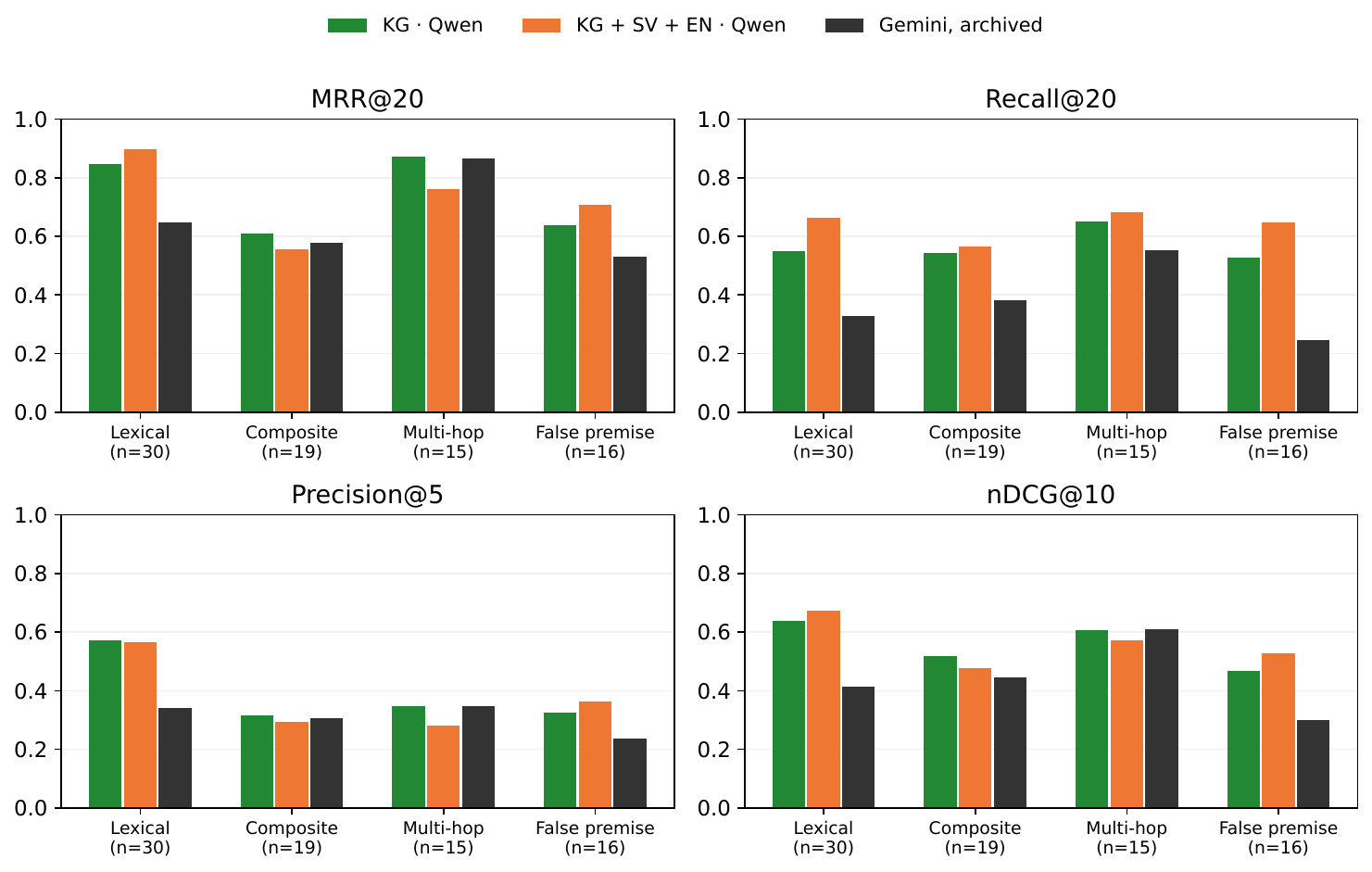}
\caption{Category-specific internal strengths against the archived Gemini reference: Lexical Understanding, Composite, Multi-hop, and False Premise. Internal bars use Qwen and show KG alone and the KG + SV + EN union. All four metrics are displayed, including the small Multi-hop nDCG loss to Gemini. Category sample sizes are shown below the axes.}\label{fig:gemini-category-gains}
\Description{Four metric panels compare Qwen KG, Qwen KG plus SV plus EN, and Gemini across Lexical, Composite, Multi-hop, and False Premise questions. Internal gains depend on category, metric, and union membership.}
\end{figure*}

\begin{figure*}[t]\centering
\includegraphics[width=\linewidth]{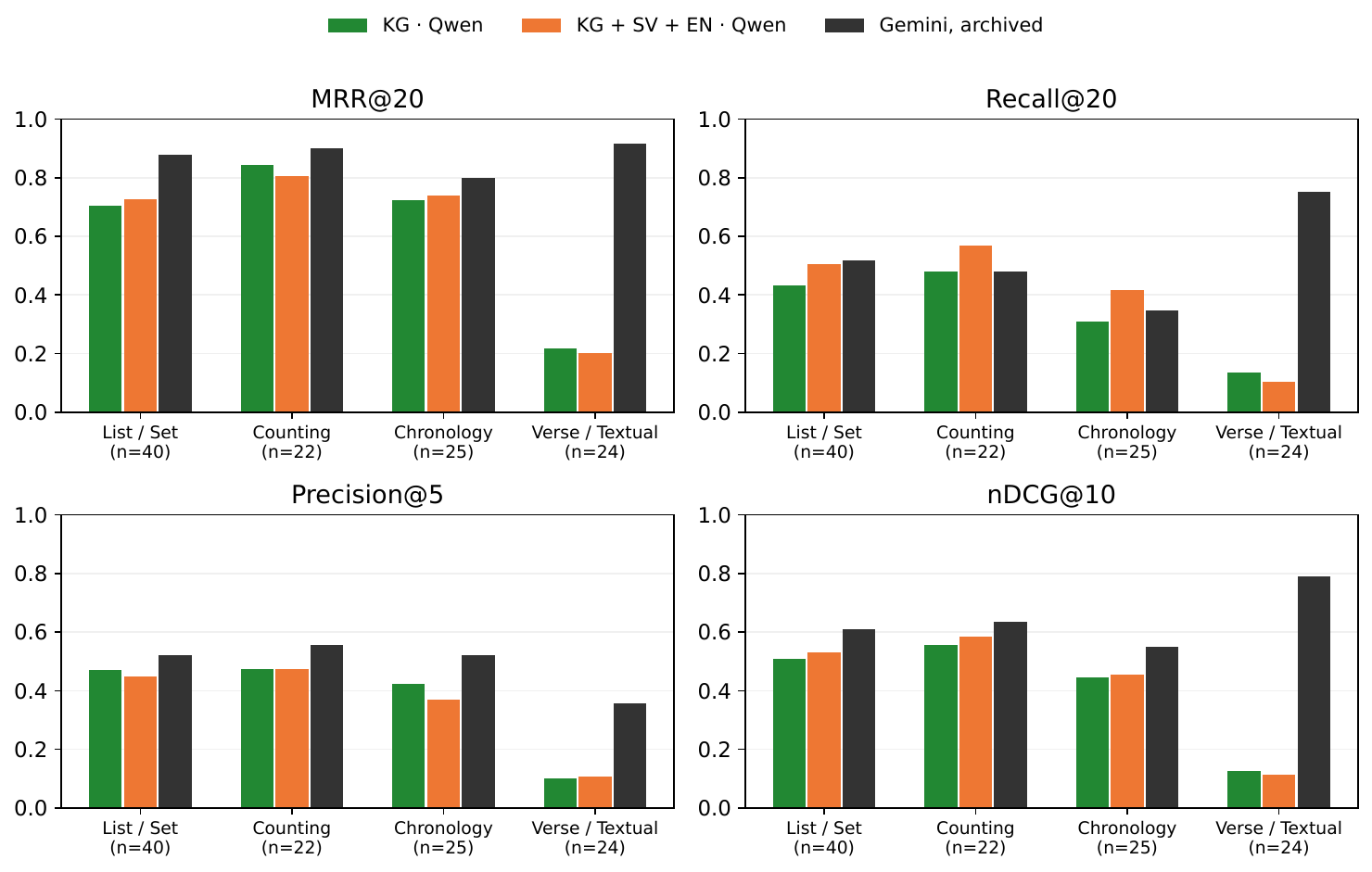}
\caption{Categories retaining substantial external-baseline advantages. Gemini leads Qwen KG and the Qwen EN union on all four metrics in List/Set and Verse/Textual Retrieval. Counting and Chronology show metric-specific trade-offs: the EN union improves recall while Gemini retains stronger first-hit ranking, precision, and nDCG.}\label{fig:gemini-category-losses}
\Description{Four metric panels compare Qwen internal systems with Gemini on List/Set, Counting, Chronology, and Verse/Textual Retrieval, showing external ranking advantages and selected internal recall gains.}
\end{figure*}

\begin{figure*}[t]\centering
\includegraphics[width=\linewidth]{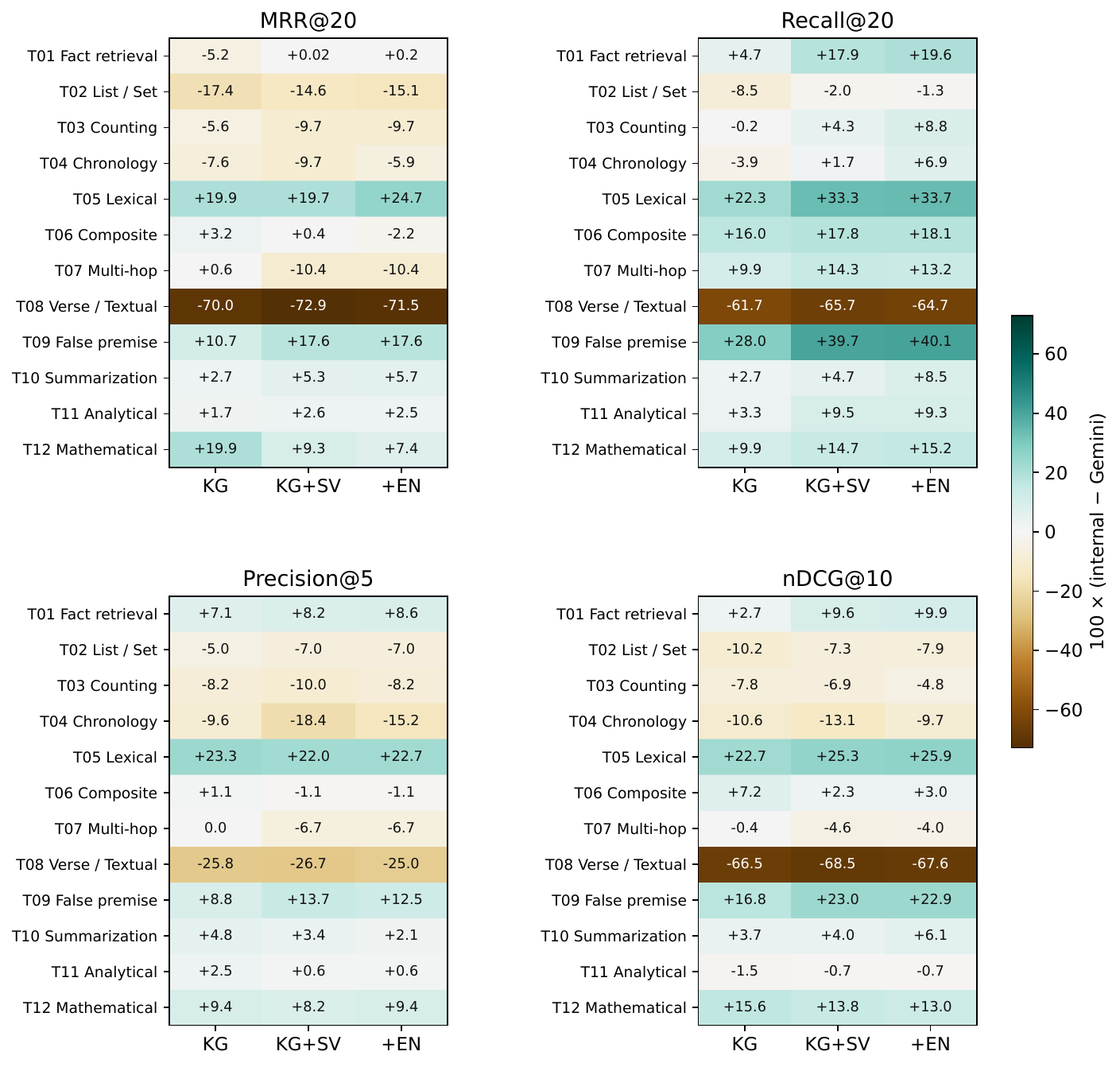}
\caption{Category-resolved Qwen internal-system scores minus Gemini, multiplied by 100. Positive cells favor the internal system; negative cells favor Gemini. Columns show KG, the fully reranked KG + SV union, and KG + SV + EN. The common 320-question cohort is fixed, but Gemini's retrieval budget is not matched. Values are score differences, not relative percentage changes.}\label{fig:gemini-category-matrix}
\Description{Four heatmaps show internal-minus-Gemini differences for MRR at 20, recall at 20, precision at 5, and nDCG at 10 across all twelve categories. Lexical, Composite, False Premise, and Mathematical questions contrast with strong external results on Verse/Textual and List/Set retrieval.}
\end{figure*}

The category-resolved comparison in Figures~\ref{fig:gemini-category-gains}--\ref{fig:gemini-category-matrix} reveals substantially different error profiles. Lexical Understanding is a clear internal-system strength: Qwen KG exceeds Gemini on MRR@20 (0.8481 versus 0.6492), Recall@20 (0.5507 versus 0.3276), Precision@5 (0.5733 versus 0.3400), and nDCG@10 (0.6397 versus 0.4129). The Qwen EN union further raises lexical MRR@20 to 0.8965 and Recall@20 to 0.6641. Qwen KG also exceeds Gemini on all four metrics in False Premise and Mathematical/Logical/Computational questions. These are chunk-ranking results; the False Premise category does not establish correct rejection or answer generation.

Gemini retains important advantages (Figure~\ref{fig:gemini-category-losses}). In Verse/Textual Retrieval, its MRR@20 is 0.9167 versus 0.2167 for Qwen KG, with higher recall, precision, and nDCG as well. Gemini also exceeds Qwen KG and the Qwen EN union on all four metrics in List/Set Retrieval. Chronology favors Gemini on MRR, precision, and nDCG, while the EN union improves recall from Gemini's 0.3486 to 0.4176. In Fact Retrieval, Qwen KG loses on MRR but gains on the other three metrics; adding EN reverses the MRR comparison narrowly (0.8273 versus 0.8258). Analytical questions similarly separate first-hit ranking from ranked-list quality: Qwen KG has higher MRR, recall, and precision, but lower nDCG than Gemini. Overall, Qwen KG exceeds Gemini in seven of twelve categories on MRR, eight on recall, seven on precision with one tie, and six on nDCG. The two full unions exceed Gemini's recall in ten categories. All differences are descriptive, and the external candidate budget is not controlled.

\subsection{Merging policy changes the ranking outcome}
\begin{figure*}[t]\centering
\includegraphics[width=\linewidth]{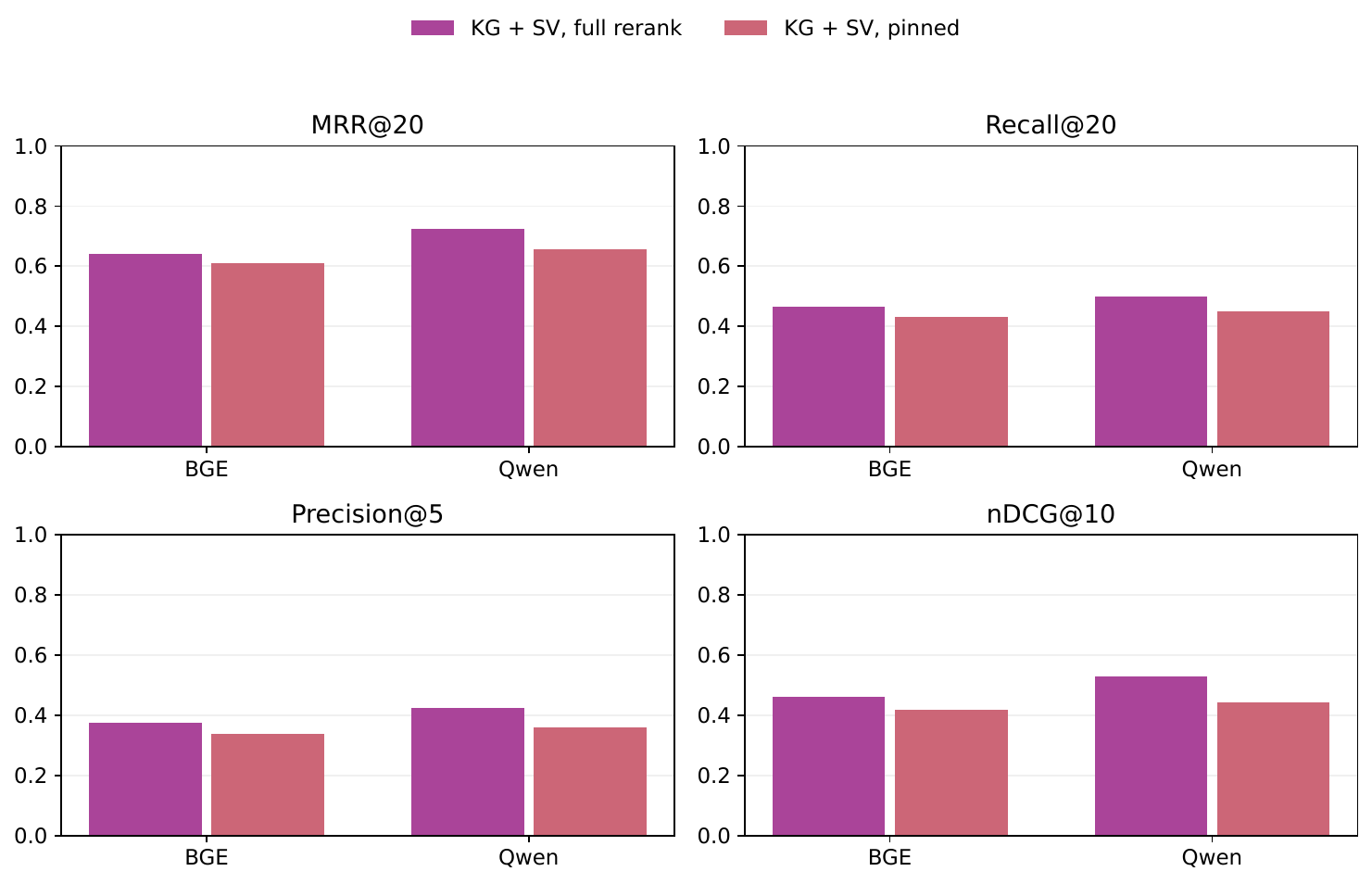}
\caption{Pinned versus fully reranked KG + SV unions on identical candidate pools. The overall result favors full reranking on all four metrics for both models; category exceptions are described in the text. ``Pinned'' uses the KG reranked order inside the fixed intersection prefix.}\label{fig:fixed-02-merging}
\Description{Pinned versus fully reranked KG + SV unions on identical candidate pools. The overall result favors full reranking on all four metrics for both models; category exceptions are described in the text. ``Pinned'' uses the KG reranked order inside the fixed intersection prefix.}
\end{figure*}

Figure~\ref{fig:fixed-02-merging} compares the full and pinned ranking policies on identical candidate pools.

\begingroup\emergencystretch=2em\relax
Fixing shared evidence at the top and reranking the entire merged pool are empirically different choices. The full and pinned unions contain exactly the same candidates for each question\slash model, so their comparison isolates ranking policy rather than candidate selection. Overall, full reranking exceeds pinning on all four reported metrics with both rerankers. With BGE, full versus pinned MRR@20 is 0.6397 versus 0.6106 and nDCG@10 is 0.4604 versus 0.4170. With Qwen, the corresponding values are 0.7245 versus 0.6553 and 0.5304 versus 0.4427; Recall@20 also increases from 0.4509 to 0.4980.\par\endgroup

\begingroup\emergencystretch=2em\relax
The overall preference for full reranking is not universal within categories. For Multi-hop, pinning produces higher MRR@20 than full reranking under both BGE (0.7704 versus 0.7035) and Qwen (0.8264 versus 0.7622). Qwen Composite MRR also slightly favors pinning (0.5883 versus 0.5826), although full reranking has higher nDCG@10. Shared-source agreement can therefore be useful as a category-specific prior, while forcing every intersection item ahead of exclusive evidence can reduce overall early ranking quality.\par\endgroup

\begin{figure*}[t]\centering
\includegraphics[width=\linewidth]{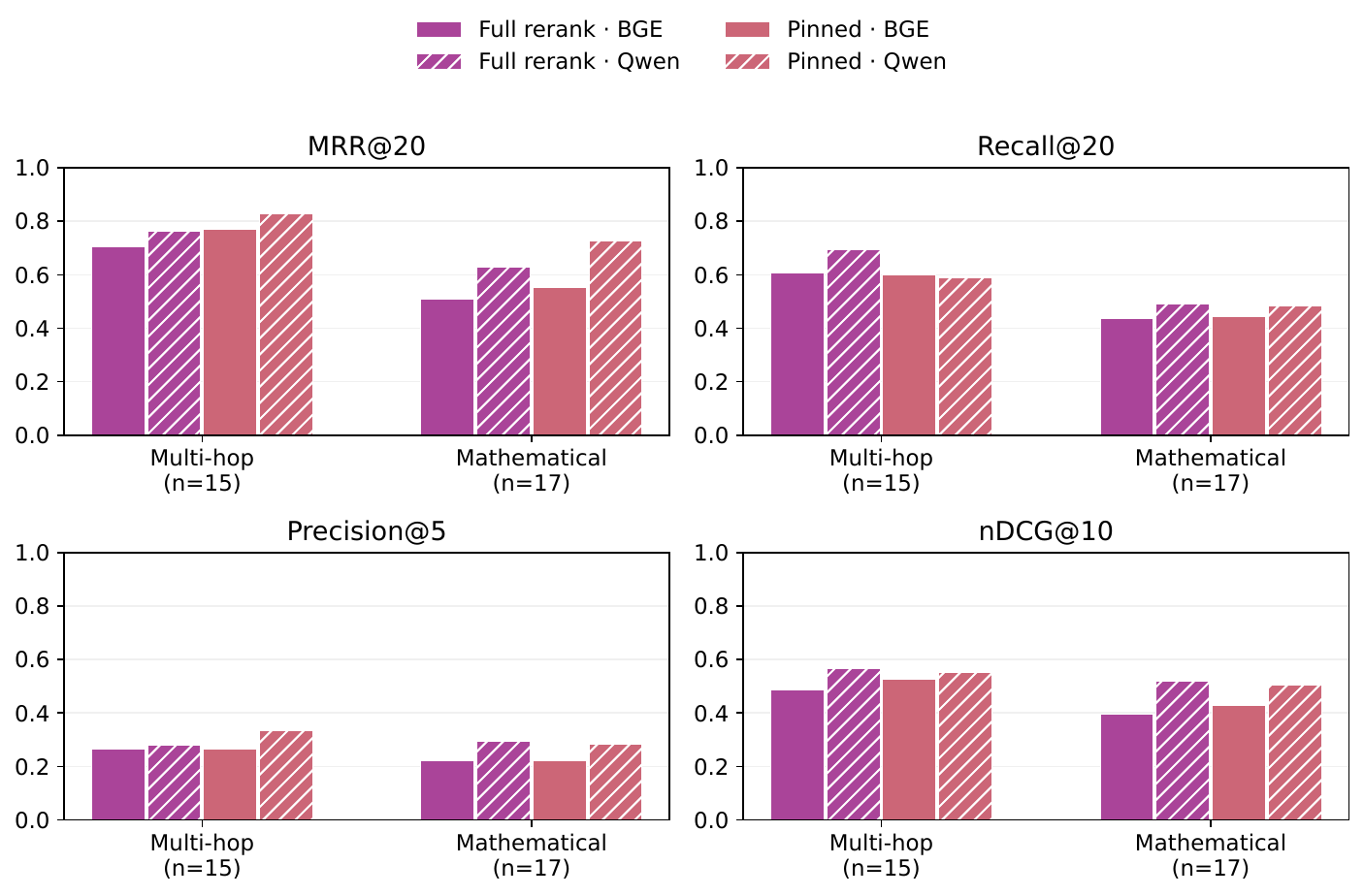}
\caption{Pinned-policy exceptions on identical candidate pools: Multi-hop and Mathematical/Logical/Computational questions. Pinning improves MRR@20 under both rerankers; other metrics can favor full reranking. Hatching denotes Qwen. The panels show all four metrics to retain the ranking--coverage trade-off.}\label{fig:merging-pinned-categories}
\Description{Four panels compare full and pinned merging under BGE and Qwen for Multi-hop and Mathematical questions. Pinning has higher MRR in both categories, while recall, precision, and nDCG have model-specific outcomes.}
\end{figure*}

\begin{figure*}[t]\centering
\includegraphics[width=0.99\linewidth]{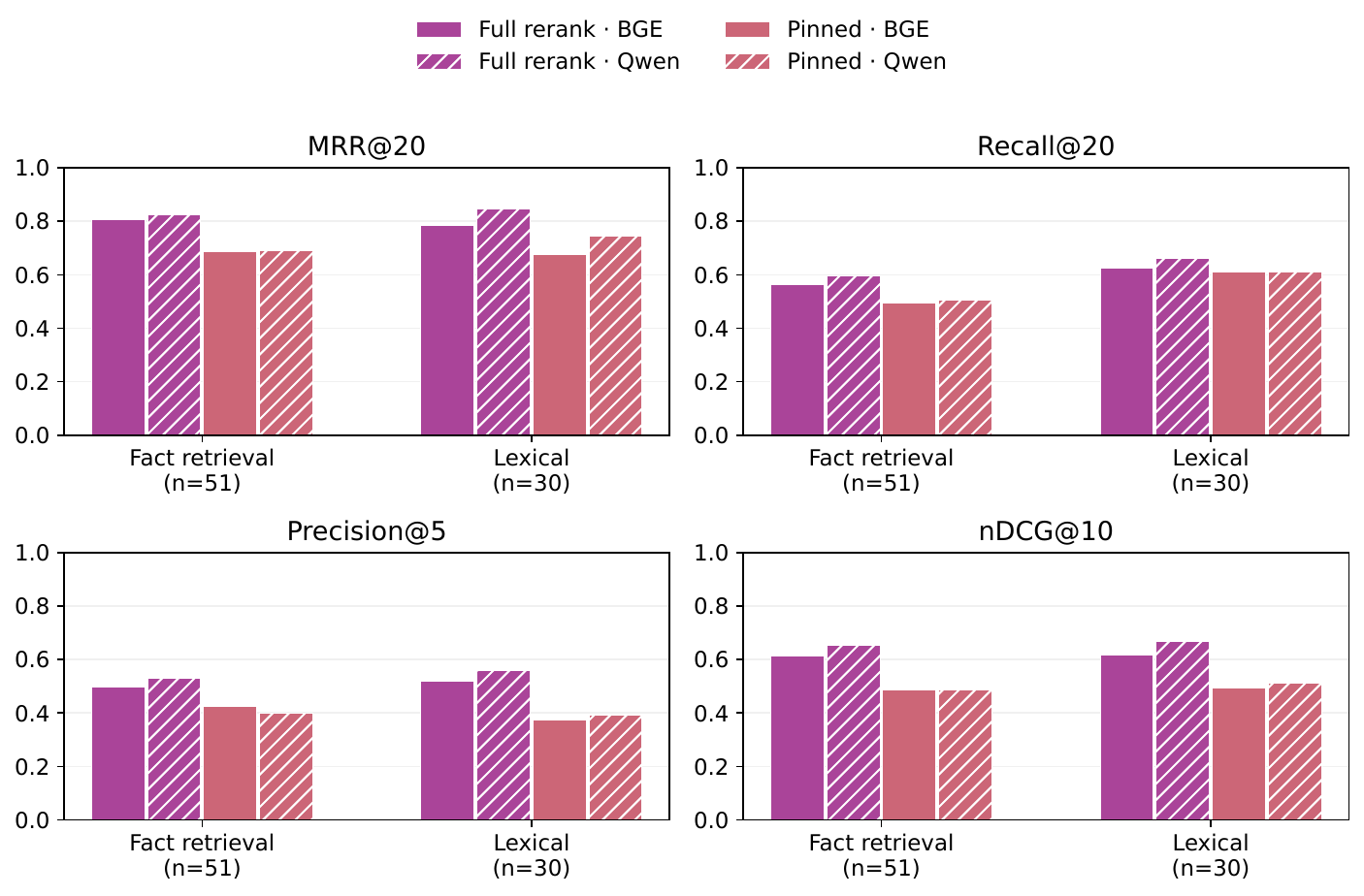}
\caption{Categories consistently favoring full reranking: Fact Retrieval and Lexical Understanding. Full reranking exceeds intersection pinning on all four metrics under both BGE and Qwen. Candidate membership and budget are identical within each policy comparison; only ordering changes.}\label{fig:merging-full-categories}
\Description{Four panels compare full and pinned merging for Fact Retrieval and Lexical Understanding under both rerankers. Full reranking has higher MRR, recall, precision, and nDCG throughout.}
\end{figure*}

Figures~\ref{fig:merging-pinned-categories} and~\ref{fig:merging-full-categories} expose these policy effects at category level. In Mathematical/Logical/Computational questions, pinning improves MRR@20 from 0.5082 to 0.5521 with BGE and from 0.6294 to 0.7255 with Qwen. However, Qwen full reranking has higher recall, precision, and nDCG in that category. Multi-hop shows a similar objective-dependent result: Qwen pinning improves MRR and Precision@5 but reduces Recall@20 from 0.6946 to 0.5902 and nDCG@10 from 0.5649 to 0.5517. In contrast, full reranking exceeds pinning on every reported metric in both Fact Retrieval and Lexical Understanding under both models. For Qwen lexical questions, MRR rises from 0.7453 to 0.8467 and Precision@5 from 0.3933 to 0.5600. Agreement-based pinning can improve the first relevant hit without improving the entire ranked list.

Complete-evidence metrics reinforce this policy dependence in Multi-hop. With BGE, pinning raises CE-MRR@20 from full reranking's 0.3361 to 0.4111 while preserving CE-Recall@20 at 8/15. With Qwen, pinning raises CE-MRR@20 from 0.3757 to 0.4556 but reduces CE-Recall@20 from 10/15 to 8/15. The preferred agreement prior therefore depends on whether the objective is earlier sufficient evidence or the completion of more questions within the cutoff.

\section{Discussion and Limitations}
\subsection{Graph Evidence and Early Relevance}
The fixed-budget results distinguish candidate quality from candidate quantity. KG's high MRR and Precision@5 indicate that its retrieved passages often place annotated, question-relevant evidence near the beginning of the list. Relations, entity neighborhoods, and source-linked provenance provide plausible mechanisms for exposing evidence that is distributed across passages or expressed through different surface forms. This interpretation is consistent with the category results, although the experiment does not isolate the causal contribution of traversal, extraction, normalization, or relationship-vector search. The metrics measure relevance against the annotated evidence sets.

KG is therefore an appropriate candidate when the application prioritizes the first useful source or a short, precise evidence list. It is not uniformly the best early-ranking system: Qwen KG + SV + EN slightly exceeds standalone KG in overall MRR and nDCG. The architecture and reranker must be evaluated together, rather than treating a graph as sufficient to guarantee superior ranking.

\subsection{Counting, Composite, and Multi-hop Workloads}
The strongest first-hit case for standalone KG arises in Counting, Composite, and Multi-hop questions, where it has higher MRR@20 than SV and either full union under both rerankers. Multi-hop provides the clearest structural motivation: a useful evidence set can require connections across entities and source passages. Composite questions aggregate several evidence requirements, while counting questions require identifying members of a relevant set. The graph's explicit organization is useful for these tasks, but a high MRR alone does not establish that every required member or supporting passage was recovered.

The category-specific benefits must consequently be weighed against completeness. Qwen unions provide higher Recall@20 than KG in all three focal categories, and the EN union improves Counting nDCG despite its lower MRR. A workload emphasizing rapid identification of a relevant relational source can justify KG; an exhaustive counting or multi-requirement workflow may instead favor a union. Composite questions are evaluated as complete multi-requirement questions rather than through an interactive dialogue with retrieval after each turn. No claim of multi-turn conversational competence follows from chunk-level retrieval scores.

\subsection{Precision and Recall Imply Different Evidence Policies}
The opposite precision--recall trends provide a practical choice between a concentrated evidence list and broader annotated coverage. Standalone KG has the highest overall Precision@5 under both rerankers, whereas the Qwen EN union raises Recall@20 from 0.4347 to 0.5134. For a reader or generator operating with a small context window, early precision and MRR may be the primary objectives. For evidence synthesis, audit, or exhaustive collection, recall and the completeness of supporting sets may matter more.

Broader coverage here means more annotated relevant passages within the evaluation cutoff, not a larger candidate-count allowance: all primary internal systems retain exactly the original per-question KG budget. A service can omit KG and use SV when its category-specific quality and latency are sufficient, but SV alone does not achieve the strongest overall recall in this experiment. The largest overall recall comes from combining KG, SV, and mapped EN evidence. The complete-evidence results in Section~\ref{sec:results-complete-evidence} make this distinction concrete: Qwen KG completes Multi-hop evidence earlier on average, whereas the EN union completes ten rather than nine of the fifteen questions within twenty results. Retrieved-set completeness and downstream answer correctness remain separate evaluation requirements.

\subsection{Architecture Depends on Hardware, Question Pool, and Affordable Latency}
The GPU study establishes a substantial practical latency trade-off. KG retrieval averaged 20.374 s versus 0.646 s for SV, an additional 19.728 s, or approximately twenty seconds per query. With passage reranking included as separately measured stage sums, the KG--SV gap remains approximately twenty seconds under either BGE or Qwen. Relational and aggregation gains must therefore justify a material response-time premium; the measured advantage on Multi-hop does not make KG the appropriate default for every query.

Hardware capacity is part of this decision. Graph generation with Qwen3-32B incurs offline extraction and maintenance work, while online KG processing and reranking require additional model residency and accelerator capacity. In the sampled deployment, KG's per-domain embedding-model caches occupied approximately 22.6 GiB of A100 memory after initialization. This is an implementation-specific observed footprint rather than a minimum requirement. The isolated KG instance was stopped after retrieval to release that cache before scoring; simultaneous serving and reranking would require its own capacity and concurrency assessment.

A deployment should therefore select among SV, KG, and unions according to the available hardware, expected question mixture, and affordable latency. SV is a strong starting point for strict interactive response limits; KG is more defensible when relational quality gains outweigh an approximately twenty-second retrieval premium on representative questions. Category-aware routing is a plausible way to concentrate that cost on suitable requests, but it was not evaluated here and must not rely on ground-truth category labels at inference time. The measured gap is a property of these shared-machine implementations, not a hardware-independent constant or a validated routing threshold.

\subsection{Reranker Quality Must Be Budgeted Separately}
Qwen generally improves quality across the internal systems, with local BGE exceptions that can matter for a particular category or metric. Its approximately 7.3-fold reranking-time increase should be interpreted in absolute as well as relative terms: moving from BGE to Qwen adds about 2.2 s per question for either source pool. This is modest relative to KG retrieval, which contributes 88.8\% of the KG+Qwen composed latency, but consequential for SV, where Qwen scoring contributes 79.6\% of the total. Reranker choice therefore has a different operational effect in the two architectures.

BGE is attractive when its category-level quality is adequate and the latency or memory budget is restrictive; Qwen is attractive when its relevance gains justify the additional scoring work. These comparisons apply to the documented precision, batch sizes, and 1,024-token passage cap. Reranking time is more strongly correlated with capped document-token volume than with candidate count, so an equal-count budget should be accompanied by token-volume and model-specific measurements. Construction, retrieval, and reranking should be budgeted separately rather than attributing the entire runtime difference to the scorer.

\subsection{Gemini Reveals Complementary Category Strengths}
The external comparison supports category- and metric-specific claims rather than a universal superiority claim. Qwen KG is stronger than the archived Gemini reference on all four reported metrics in Lexical Understanding, Composite, False Premise, and Mathematical/Logical/Computational questions. Multi-hop is closely matched on MRR and precision, with a KG recall advantage and a small Gemini nDCG advantage. These results indicate useful internal retrieval strengths, while the large Verse/Textual gap and Gemini's List/Set and Counting ranking advantages identify important weaknesses.

Unions frequently improve coverage against the external reference without reversing every ordering disadvantage. Both Qwen full unions exceed Gemini's Recall@20 in ten categories, yet Gemini retains the highest overall MRR and nDCG. Consequently, the intended evidence objective and category distribution should determine which comparison is decisive. Gemini's archived output depth and retrieval budget differ from the internal candidate budgets, and its runtime was not measured in the paired timing study. Neither an equal-compute superiority claim nor a Gemini latency comparison is supported.

\subsection{Intersection Agreement Is a Ranking Prior, Not a Relevance Guarantee}
The merging-policy comparison holds candidate membership fixed and changes ordering, showing that source agreement has conditional value. Pinning intersection evidence improves MRR in Multi-hop and Mathematical questions under both models, but can sacrifice deeper recall or ranked-list quality. Full reranking is the stronger general default in the observed aggregate and consistently improves every displayed metric in Fact Retrieval and Lexical Understanding. Shared IDs can therefore identify useful evidence without guaranteeing that every shared passage belongs ahead of every exclusive passage.

The two policies serve different hypotheses: pinning imposes a fixed agreement prior, while full reranking allows question-conditioned passage scores to override it. A category-specific policy could be investigated on held-out data, but the present category exceptions do not establish a validated selector. When the KG-reranked order is already available, the pinned policy reranks only the non-intersection remainder. However, this study does not provide a controlled timing comparison between the two merging policies; no policy-specific latency saving is claimed.

\subsection{Extraction, Identity, and Provenance Risks}
The pipeline relies on model-generated entities and relations, prompt-based ontology adherence, and heuristic entity merging. Errors at any of these stages can produce plausible but unsupported graph connections. No separate expert-labeled triple-quality experiment is reported, and the evaluated construction run omitted optional relation validation. Retrieval gains therefore must not be recast as a measured improvement in graph factuality.

Aliases and summary similarity help reconcile name variants but can conflate different senses. Conversely, conservative merging can fragment a single entity. Hand-authored ambiguity rules address selected cases rather than solving identity resolution generally. Endpoint-based relation merging and fallback chunk mapping can also lose or broaden evidence associations. Scoring original passages assesses candidate relevance after graph discovery but cannot recover absent evidence, certify extracted triples, or guarantee complete support.

\subsection{Evaluation and Reproducibility Boundaries}
Quality comparisons use the exact 320-question cohort and frozen Dhara-Bench v3 ground truth, with per-question candidate counts matched across the primary internal systems. The benchmark contains 325 questions overall; the five absent KG predictions are excluded rather than imputed. Gemini has the same evaluation questions but an unmatched retrieval budget. Precision@5 retains denominator five, recall depends on the breadth of the annotated evidence sets, and binary nDCG does not distinguish degrees of relevance. Complete-evidence metrics use only the 34 group-annotated questions (19 Composite/Multi-turn and 15 Multi-hop), all of which are evaluable. All fixed-budget, complete-evidence, and category-resolved comparisons are descriptive.

The timing study uses a deliberately category-balanced 24-question sample rather than all 320 questions. Two passes expose repeat effects, but two unique questions per category do not establish stable category-specific latency estimates. GPU placement was explicit for both retrieval systems, while their serving stacks, device residency, database access, and background load differed. The reported totals sum separately measured retrieval and reranking stages; they do not include answer generation, concurrent serving, or a controlled end-to-end union request. The live databases were not frozen to the historical quality export.

The retrieval heuristics were developed within the project benchmark. We do not report an independent held-out evaluation or a blinded annotation audit. The results therefore characterize the implemented pipeline against the recorded evidence sets, not label-independent or out-of-domain superiority. Equal candidate counts do not equalize token work, construction cost, or total compute, and the comparisons do not isolate each graph component causally. Complete-evidence retrieval is assessed separately from ordinary relevance; neither type of retrieval score alone establishes downstream answer faithfulness.

Exact reconstruction of the evaluated graph requires the source texts, extraction and serving configurations, provenance mappings, and a database snapshot tied to the extraction run. The current deployment census and timing measurements do not establish an identical historical graph snapshot. This limits exact reconstruction and the interpretation of the timing results as a quality--latency measurement for the historical evaluation. The graph and KG-specific implementation are not publicly released with this manuscript.

\section{Conclusion}
We presented an ontology-guided KG construction and evidence-retrieval pipeline for Indian Knowledge Systems. On the common 320-question \dhara{} Bench cohort, fixed per-question candidate budgets distinguish early relevance from broader evidence coverage. Standalone KG has the highest Precision@5 under both rerankers and the highest MRR@20 in Counting, Composite, and Multi-hop among the evaluated internal configurations. With Qwen, KG + SV + EN provides the highest overall MRR@20 (0.7306), Recall@20 (0.5134), and nDCG@10 (0.5376), while KG retains higher Precision@5 (0.4419). Complete-evidence results show the same distinction: Qwen KG attains Multi-hop CE-MRR@20 of 0.4651, whereas both full unions complete ten of fifteen questions within twenty results compared with nine for KG. The reviewed Gemini reference remains stronger on overall MRR and nDCG, while internal systems offer advantages in selected categories and evidence coverage. All comparisons are descriptive. A separate deployment study measures a substantial retrieval-time premium for KG, motivating selection according to the evidence objective, question mixture, and available resources. Component-level causal attribution, expert assessment of graph factuality, and held-out generalization remain open questions.

\section*{Data and Code Availability}
The corpus and evaluation code for \emph{\dharabenchtitle}~\citep{dharabench} are available at \url{https://github.com/Avinash2236/Dhara-Bench}. The companion retrieval study is available as arXiv:2611.11197~\citep{dhara}. No public release of the KG-specific code, evaluated graph, or experiment bundles accompanies this manuscript.

\section*{Disclosure of AI Assistance}
OpenAI Codex with GPT-6.1 Sol was used to assist manuscript drafting and editing, and Astra was used for final edits. These writing tools are distinct from the models used in the reported extraction and retrieval experiments. The authors are responsible for the manuscript's claims, references, and reported results.

\FloatBarrier
\section*{Acknowledgements}
An early proof of concept for this work was supported by the Department of Science and Technology (DST), Government of India, through its Science and Heritage Research Initiative (SHRI), under grant no. DST/TDT/SHRI-34/2021. This work was supported by a Corporate Social Responsibility (CSR) grant from Perfios Software Solutions Private Limited to HTSR Institute. One of the joint first authors was supported through the Adani Indology Fellowship, funded by Adani Enterprises Limited.

We thank Adiyen Technologies Private Limited for assistance with data preparation.

\bibliographystyle{ACM-Reference-Format}
\bibliography{references}

@article{qwen3,
 author={Yang, An and others},
 title={{Qwen3} Technical Report},
 journal={arXiv preprint arXiv:2505.09388},year={2025},
 url={https://arxiv.org/abs/2505.09388}}

@article{qwenembedding,
 author={Zhang, Yanzhao and Li, Mingxin and Long, Dingkun and Zhang, Xin and Lin, Huan and Yang, Baosong and Xie, Pengjun and Yang, An and Liu, Dayiheng and Lin, Junyang and Huang, Fei and Zhou, Jingren},
 title={{Qwen3 Embedding}: Advancing Text Embedding and Reranking Through Foundation Models},
 journal={arXiv preprint arXiv:2506.05176},year={2025},
 url={https://arxiv.org/abs/2506.05176}}

@article{bgem3,
 author={Chen, Jianlv and Xiao, Shitao and Zhang, Peitian and Luo, Kun and Lian, Defu and Liu, Zheng},
 title={{BGE M3-Embedding}: Multi-Lingual, Multi-Functionality, Multi-Granularity Text Embeddings Through Self-Knowledge Distillation},
 journal={arXiv preprint arXiv:2402.03216},year={2024},
 url={https://arxiv.org/abs/2402.03216v4}}

@misc{bgereranker,
 author={{Beijing Academy of Artificial Intelligence}},
 title={{BAAI/bge-reranker-v2-m3}: Model Card},
 howpublished={Hugging Face},year={n.d.},
 url={https://huggingface.co/BAAI/bge-reranker-v2-m3},
 note={Accessed 6 October 2026}}

@article{graphrag,
 author={Edge, Darren and Trinh, Ha and Cheng, Newman and Bradley, Joshua and Chao, Alex and Mody, Apurva and Truitt, Steven and Larson, Jonathan},
 title={From Local to Global: A Graph {RAG} Approach to Query-Focused Summarization},
 journal={arXiv preprint arXiv:2404.16130},year={2024},
 url={https://arxiv.org/abs/2404.16130}}

@article{lightrag,
 author={Guo, Zirui and Xia, Lianghao and Yu, Yanhua and Ao, Tu and Huang, Chao},
 title={{LightRAG}: Simple and Fast Retrieval-Augmented Generation},
 journal={arXiv preprint arXiv:2410.05779},year={2024},
 url={https://arxiv.org/abs/2410.05779}}

@article{colbert,
 author={Khattab, Omar and Zaharia, Matei},
 title={{ColBERT}: Efficient and Effective Passage Search via Contextualized Late Interaction over {BERT}},
 journal={arXiv preprint arXiv:2004.12832},year={2020},
 url={https://arxiv.org/abs/2004.12832}}

@inproceedings{rrf,
 author={Cormack, Gordon V. and Clarke, Charles L. A. and Buettcher, Stefan},
 title={Reciprocal Rank Fusion Outperforms {Condorcet} and Individual Rank Learning Methods},
 booktitle={Proceedings of SIGIR},pages={758--759},year={2009},
 doi={10.1145/1571941.1572114},
 url={https://plg.uwaterloo.ca/~gvcormac/cormacksigir09-rrf.pdf}}

@article{rag,
 author={Lewis, Patrick and others},
 title={Retrieval-Augmented Generation for Knowledge-Intensive {NLP} Tasks},
 journal={Advances in Neural Information Processing Systems},volume={33},year={2020},
 url={https://arxiv.org/abs/2005.11401}}

@article{dhara,
 author={Nagendra, Avinash Holenarasipur and Prasad, Anand Kaniyaar Sriranga and Vishwakarma, Rahul and Gavit, Trupti and Tipnis, Anish Amol and Rangayyan, Yashaswini Mandayam and Appajanna, Vinod and Sridharan, Kousik Sarathy and Raghavan, Mohan},
 title={{Dhara: Evidence-Oriented Retrieval for Indian Knowledge Systems}},
 journal={arXiv preprint arXiv:2611.11197},
 year={2026},
 eprint={2611.11197},
 archivePrefix={arXiv},
 primaryClass={cs.IR},
 doi={10.48550/arXiv.2611.11197},
 url={https://arxiv.org/abs/2611.11197}}

@article{dharabench,
 author={Nagendra, Avinash Holenarasipur and Prasad, Anand Kaniyaar Sriranga and Sridharan, Kousik Sarathy and Raghavan, Mohan},
 title={{Dhara Bench: An Evidence-Oriented Retrieval Benchmark for Indian Knowledge Systems}},
 journal={arXiv preprint arXiv:2611.03196},
 year={2026},
 eprint={2611.03196},
 archivePrefix={arXiv},
 primaryClass={cs.IR},
 doi={10.48550/arXiv.2611.03196},
 url={https://arxiv.org/abs/2611.03196}}

@book{elements_iks,
 author={Raghavan, Mohan and {Harsha Simha M. S.} and Ramaswamy, C. R.},
 title={{Elements of Indic Knowledge Systems \& Heritage}},
 publisher={HTSR Institute},
 year={2023},
 isbn={9788196417604},
 url={https://sites.google.com/view/htsr-institute/publications/books/elements-of-iksh}}
\end{document}